\documentclass[a4paper,fleqn]{cas-sc}

\usepackage[authoryear,longnamesfirst]{natbib}
\usepackage{caption}
\usepackage{subcaption}

\def\tsc#1{\csdef{#1}{\textsc{\lowercase{#1}}\xspace}}
\tsc{WGM}
\tsc{QE}
\tsc{EP}
\tsc{PMS}
\tsc{BEC}
\tsc{DE}

\begin{document}
\let\WriteBookmarks\relax
\def\floatpagepagefraction{1}
\def\textpagefraction{.001}

\shorttitle{Text-to-Image Representativity Fairness Evaluation Framework}

\shortauthors{Yamani and Baslyman}

\title [mode = title]{Text-to-Image Representativity Fairness Evaluation Framework}                      



%
\author[1]{Asma Yamani}[orcid=0000-0002-6277-8972]



\ead{g201906630@kfupm.edu.sa}


\credit{Conceptualization, Methodology, Software, Formal analysis, Data Curation, Writing - Original Draft}

\affiliation[1]{organization={King Fahd University of Petroleum and Minerals},
    city={Dhahran},
    postcode={31261}, 
    country={Saudi Arabia}}

\author[1]{Malak Baslyman}[orcid=0000-0003-4002-4480]
\cormark[1]
\ead{malak.baslyman@kfupm.edu.sa}

\credit{Conceptualization, Validation, Writing - Review \& Editing, Supervision }




\cortext[cor2]{Principal corresponding author}



\begin{abstract}
Text-to-Image generative systems are progressing rapidly to be a source of advertisement and media and could soon serve as image searches or artists.
However, there is a significant concern about the representativity bias these models embody and how these biases can propagate in the social fabric after finetuning them. Therefore, continuously monitoring and evaluating these models for fairness is important.
To address this issue, we propose Text-to-Image (TTI) Representativity Fairness Evaluation Framework.
In this framework, we evaluate three aspects of a TTI system; diversity, inclusion, and quality. For each aspect, human-based and model-based approaches are proposed and evaluated for their ability to capture the bias and whether they can  substitute each other. The framework starts by suggesting the prompts for generating the images for the evaluation based on the context and the sensitive attributes under study. Then the three aspects are evaluated using the proposed approaches. Based on the evaluation, a decision is made regarding the representativity bias within the TTI system.
The evaluation of our framework on Stable Diffusion shows that the framework can effectively capture the bias in TTI systems. The results also confirm that our proposed model based-approaches can substitute human-based approaches in three out of four components with high correlation, which could potentially reduce costs and automate the process. The study suggests that continual learning of the model on more inclusive data across disadvantaged minorities such as Indians and Middle Easterners is essential to mitigate current stereotyping and lack of inclusiveness.
\end{abstract}






\maketitle

\section{Introduction}
Generative models learn to estimate the training data distribution and can be used for various regression, classification, and generation problems. In deep generative models, a high-dimensional probability distribution is learned by training a neural network with multiple hidden layers using many samples~\cite{https://doi.org/10.48550/arxiv.2103.05180}. Deep generative models hold a sophisticated architecture and require abundant data to train on. This data is usually obtained through crawling the internet~\cite{https://doi.org/10.48550/arxiv.2208.00406,Holstein_2019}. The output of a deep generative model can be visual, textual, or any other data type. Diffusion models have emerged in the last three years as the new state-of-the-art family of deep generative models~\cite{yang2023diffusion}. Diffusion models concerning text-to-image generation include models such as Stable Diffusion~\cite{nokey} by StabilityAI, Imagen \cite{saharia2022photorealistic} by Google, and DALL·E \cite{ramesh2022hierarchical} by OpenAI. Diffusion models have achieved tremendous success in the last few years and gained much media attention as demos are released to the public. Some are also open-sourced so that developers can build on top of them in a step called fine-tuning. Fine-tuning enables the building of high-performing models for various applications with minimal data and training costs. \par
Aside from the success, several fairness-related concerns have surfaced due to the biases these models encompass. As with other ML models, these biases include under-representation, misrepresentation, and over-representation \cite{Holstein_2019}. The concern grows larger as these models can be used as a base for different downstream tasks and may lead to the reinforcement of the biases indirectly to other applications \cite{bommasani2021opportunities, https://doi.org/10.48550/arxiv.2211.03759} resulting in multiple types of harms such as quality-of-service, stereotyping, denigration, or under-representation \cite{10.1145/3386296.3386298,friedrich2023fair,luccioni2023stable}. \par
There is a large body of work on the methods and tools that are developed to audit discriminative models (classification and regression models) and help in detecting their underlying biases~\cite{DBLP:journals/corr/abs-2012-09951, FairnessEvaluation,jaccard_2022,wexler2019if}. Most of the work is based on statistical fairness metrics such as Statistical (Demographic) parity, Equal opportunity, and Equalized odds requiring a ground truth to uncover the biases~\cite{https://doi.org/10.48550/arxiv.2207.07068}. Therefore, such approaches are more suitable for detecting allocation and quality-of-service harms in classification and regression problems. However, covering the types of harms that generative models can present (e.g., stereotyping, denigration, or under-representation harms) is beyond their scope~\cite{DBLP:journals/corr/abs-2012-09951,Holstein_2019,https://doi.org/10.48550/arxiv.2211.03759}. The work on fairness in generative models is an emerging topic. Recent work focused on enforcing diversity during the model training phase~\cite{xu2018fairgan,choi2020fair}, after deployment by prompt engineering~\cite{friedrich2023fair}, and through analytical evaluation~\cite{friedrich2023fair,luccioni2023stable}. As fairness of a generative model extends beyond diversity, there is a need to investigate multiple aspects that contribute to obtaining a Text-to-Image (TTI) system free from representativity biases.\par

 To conduct our research, we formulate the following three research questions (RQ):
\begin{itemize}
    \item \textbf{RQ1: What are the existing techniques to evaluate  TTI systems or similar systems or models for fairness?} 
    \item \textbf{ RQ2: To what extent can human-based and model-based evaluation approaches capture the bias in the different aspects of representativity fairness of TTI system?}
    \item \textbf{RQ3: To what extent can ML models substitute for human evaluation detection of under-representation and harms in text-to-image generative models?} 
\end{itemize}

To address the above questions, we propose a framework to evaluate the representativity fairness of TTI systems. We first survey the literature on evaluating image generative models, TTI systems, and image search to determine the main concepts addressed to capture representativity bias. Based on the literature, we propose treating the output of the TTI system as a subset selection problem and extending the use of the concepts of diversity and inclusion from~\cite{10.1145/3375627.3375832} to evaluate TTI systems for fairness. We also adapt statistical parity to encapsulate the system's behavior in terms of inclusion and quality when conditioned on different members of the sensitive attribute. For each concept, we provide a human-based operationalization approach and model-based alternatives. Providing both approaches is necessary as human-based approaches capture the true distribution of the context under study when using a representative sample; however, they are resource intensive and require a large representative sample that can be hard to acquire in some cases. On the other hand, model-based approaches require fewer resources but may have their biases and inaccuracies propagate through the evaluation leading to less reliable results.  We evaluate the proposed approaches in terms of their ability to capture the representativity bias in Stable Diffusion~\cite{nokey} and determine if the approaches can substitute for each other by measuring the correlation between them. Finally, we discuss any limitations and biases in the proposed approaches. 

This paper is structured as follows: in Section \ref{sec:rw}, we present some related work upon which we will build our proposed approach; in Section \ref{sec:pre}, we lay the notation used in the paper and recall some background knowledge used in our work including fairness measures and TTI systems; in Section \ref{sec:proposed_approach}, we describe in detail the proposed framework; Section \ref{sec:exp} is dedicated to the experimental analysis that has been conducted to evaluate the framework; in Section~\ref{sec:des} we discuss the results and some limitations; finally, Section \ref{sec:conc} concludes the paper.
\section{Related Work}
\label{sec:rw}

This section presents some related work on algorithmic fairness and answers for the first research question on the existing techniques to evaluate image search and image generative models for fairness. 

\subsection{Algorithmic fairness}
Finding a formal definition of fairness is a subject under debate, but it can be viewed as an antonym of discrimination~\cite{CORNACCHIA2023103224}. The EU Charter of Fundamental Rights defines the non-discrimination requirements as: "any discrimination based on any ground such as sex, race, color, ethnic or social origin, genetic features, language, religion or belief, political or any other opinion, membership of a national minority, property, birth, disability, age or sexual orientation shall be prohibited" \footnote{https://fra.europa.eu/en/eu- charter/article/21-non-discrimination}. It can also be viewed as an antonym of injustice, where injustice is "systematic and unfair discrimination or prejudice of certain individuals or groups of individuals in favor of others" ~\cite{10.1145/3512728}. \emph{fairness} is a broad concept that can vary based on the context. An informal definition that can be used is "Any case where AI/ML systems perform differently for different groups in ways that may be considered undesirable."~\cite{Holstein_2019}. With the recent emphasis on the harms of bias in AI systems, a growing body of research is focusing on the detection and mitigation of offering tools, statistical approaches, metrics, and datasets used for bias experiments~\cite{Pagano2023}.\par
People with marginalized demographic attributes are often discriminated against by AI systems. This includes discrimination based on gender~\cite{RODGER2004529,Ulloa2022, pmlr-v81-buolamwini18a,nytimesFacialRecognition}, race~\cite{pmlr-v81-buolamwini18a,nytimesFacialRecognition}, age~\cite{Park2021} socioeconomic status~\cite{facebookAssessBias}, and disability status~\cite{Guo2020}.
Several statistical metrics are used in the AI Ethics research body to measure such occurrences. The selection of the metric should be done carefully according to the problem and context, as it is mathematically infeasible to satisfy all of them~\cite{kleinberg2016inherent}. Such metrics include statistical (or demographic) parity (each member of a sensitive attribute (e.g., race) should take the positive outcome at equal rates ~\cite{tsintzou2018bias}), Equalized odds (each member of a sensitive attribute should have equal true positives and false positives rates~\cite{awasthi2020equalized}), and Counterfactual fairness (an outcome for an individual is fair if it is the same in both the actual world and a counterfactual world in which the individual belongs to a different member of the sensitive attribute~\cite{russell2017worlds}). As model discrimination in many cases is due to issues in the data used to train the models~\footnote{https://ainowinstitute.org/publication/disabilitybiasai-2019}, social-minded measures of data quality were proposed to evaluate the effect of data, algorithms, and systems in society~\cite{Pitoura2020}. Three measures were included in this set: diversity, which ensures that all relevant aspects are represented; lack of bias, described as processing data without unjustifiable concentration on a particular side; and fairness, defined in this work as non-discriminating treatment of data and people. Furthermore, the concepts of diversity and inclusion were adopted as a necessary evaluation measure in the subset selection problem \cite{10.1145/3375627.3375832}. This problem manifests in many applications that use AI models, such as search and recommendation systems. Diversity was defined as "Variety in the representation of individuals in an instance or set of instances, with respect to sociopolitical power differentials," ensuring group fairness. In the same study, inclusion was defined as "Representation of an individual user within an instance or a set of instances." Thus, greater inclusion indicates better alignment between a user to the system and the options relevant to them in the subset, ensuring individual fairness~\cite{10.1145/3375627.3375832}.

\subsection{Bias in the context of image search}
One of the earliest studies concerning bias in search emphasized that a fair search should retrieve a selection that neither exaggerates nor marginalize any particular set of items in the database in response to a set of queries~\cite{MOWSHOWITZ2002141}. Moreover, in search, ranking and not only presence should be considered during the evaluation. Therefore, statistical parity definition was modified in which a ranking scheme exhibits statistical parity if membership in a protected group does not influence an item's position in the output~\cite{10.1145/3085504.3085526}. Moreover, metrics that capture the statistical parity measurements were modified to capture the new definition. For example, Kullback-Leibler (KL) was modified to the Normalized discounted KL-divergence (rKL) to compute the expectation of the difference between protected group membership at $top-i^{th}$ appearances and the membership in the overall population~\cite{10.1145/3085504.3085526}. Bias in web search is not only exemplified in the content in the web and retrieval and ranking algorithms; it also extends to the nature of the query string, as people from different age groups and gender tend to use different levels of description and when querying, which leds to different search results. This was observed when searching for medical conditions by different genders and age groups~\cite{Yom-Tov2019-eb}.\par
When it comes to image search in particular, repeatedly, women and people of color are underrepresented~\cite{Ulloa2022,10.1145/3449100,10.1145/2702123.2702520}. This is found to heavily affect the perception of the representation of occupations under study to the real world in participants~\cite{10.1145/3449100}. Moreover, it affects the self-reported level of interest in each occupation, perception of its inclusivity, and expectations of whether they would be valued in that field~\cite{10.1145/3449100}. In addition, in image search, there is also the issue of face-ism. Face-ism relates to biases originating from the face-to-body ratio in retrieved images~\cite{Ulloa2022}. Men are presented usually with a higher face-to-body ratio leading to a greater emphasis on agentic traits, intelligence, ambition, assertiveness, and dominance. In contrast, women are presented with a higher proportion of their body visible strongly linked to sexism~\cite{Ulloa2022}.

\subsection{Bias in the context of generated Images}
As TTI systems are increasing in popularity with the power and ease to be used to produce advertisements and other media, it is important to capture and mitigate biases to avoid them being propagating to downstream tasks. Post-processing techniques can be done by prompt engineering \cite{bansal2022texttoimage} or by filtering unwanted output~\cite{schramowski2023safe}. Semantic Guidance (SEGA)~\cite{brack2023Sega} is proposed to guide the generation during prompting. In~\cite{friedrich2023fair}, SEGA pipeline~\footnote{https://github.com/ml-research/semantic-image-editing} was to mitigate gender bias in occupation-related Stable Diffusion generated images by adding + "female person" -"male person" for occupations biases in favor of male persons and steers the generation to increase fairness. In~\cite{friedrich2023fair}, it was also emphasized how the different components of a TTI system contribute to the bias, including the dataset (LAION-5B in the case of Stable Diffusion) and the encoder (CLIP in the case of Stable Diffusion). As for evaluating diversity, an aspect of fairness, in TTI systems,~\cite{luccioni2023stable} used a dense embedding that projects the images into a multidimensional vector space that is clustered and mapped to gender and race markers. Based on the entropy of the clusters, the higher the entropy, the more diverse the output of the model. The study also contributed to the field by providing a set of interactive tools to allow for a more in-depth exploration of generated images~\cite{luccioni2023stable}. \par

\subsection{Contextualizing this work}
This work contributes to the evaluation approaches of TTI systems. It proposes and evaluates a Text-to-Image Representativity Fairness Evaluation Framework. This framework extends the evaluation of TTI systems for representativity fairness to include multi-class statistical parity inclusion and quality fairness, in addition to diversity that was discussed in the literature. For each aspect of the evaluation, both human-based approaches and model-based approaches are introduced to accommodate different contexts and resources available. This work also presents a working example that illustrates the operational aspect of the different approaches and their alternatives. 
\section{Preliminaries}
\label{sec:pre}
This section introduces preliminary concepts, notations, and definitions used throughout the paper.

\subsection{Notation}
We assume $Q$ is a fair distribution, selected in this study to be uniform on the discrete sample space, but could be otherwise if implied in the paper or if per the audited problem requirements. $P$ is the distribution achieved by the generation model on the discrete sample space. $\hat{Y}$ is a predictor model, and $\hat{G}$  is a generative model, where $y$ and $g$ is the output of the models, respectively. $A$ is the sensitive attribute under study, and $a$ is an instance of this sensitive attribute space. $X$ refers to the variables resulting in the instance's output excluding $A$, and $q$ refers to the query used to generate the images with different outputs using the desired attributes $q_{x, a}$. $Inc_{F}$ refers to the set of features under study for inclusion of representativity attributes which may include a subset of $X$ or other attributes not necessarily part of producing the output of $\hat{Y}$.

\subsection{Fairness Measures}
Analogous to image search, a TTI system achieves statistical parity if the membership of a sensitive group does not influence the probability of being the outcome of the generation~\cite{yang2016measuring}, hence a model with high diversity. It should also maintain the same quality of generated samples and inclusivity to users regardless of their association with a membership of a sensitive group~\cite{10.1145/3375627.3375832}. The following presents three statistical metrics that capture this intuition.

\textbf{Kullback-Leibler divergence (KL)} measures the expectation of the logarithmic difference between two discrete probability distributions $P$ and $Q$.

\begin{equation}
    D_{KL} \left( P \middle\| Q \right) = \sum_{a} P(a)log(\frac{P(a)}{Q(a)}),
\end{equation}
\textbf{Total Variational Distance (TVD)} measures the distance between two discrete probability distributions $P$ and $Q$ in the half L-Norm.

\begin{equation}
    TVD= \frac{1}{2}||P(A)-Q(A)||,
\end{equation}

\textbf{Multi-class statistical parity} statistical parity initially targets problems such as binary classification. It states that a predictor $\hat{Y}$ of $Y$ achieves statistical parity if 
\begin{equation}
P(\hat{Y}|A=1)=P(\hat{Y}|A=0),
\end{equation}

Likewise, a TTI system $\hat{G}$ holds multi-class statistical parity for inclusivity and quality  if the inclusivity and quality scores of the generated image  $g =\hat{G}(q_{a})$ equates to the expected value of the scores across the sample space of $A$. Relaxing the definition and applying it to multi-class problems by comparing , we define the multi-class statistical parity for any desired attribute measured by $f_{score}$ as: 

\begin{equation}
\forall a\in A |P(f_{score}(\hat{G}(q)|A=a) - \mathop{\mathbb{E}}_{a\in A}(f_{score}\hat{G}(q)|A=a)| \leq \epsilon,
\end{equation}

\subsection{Text-to-Image (TTI) Generative Diffusion Models}
\begin{figure}
    \centering
    \includegraphics[width=\linewidth]{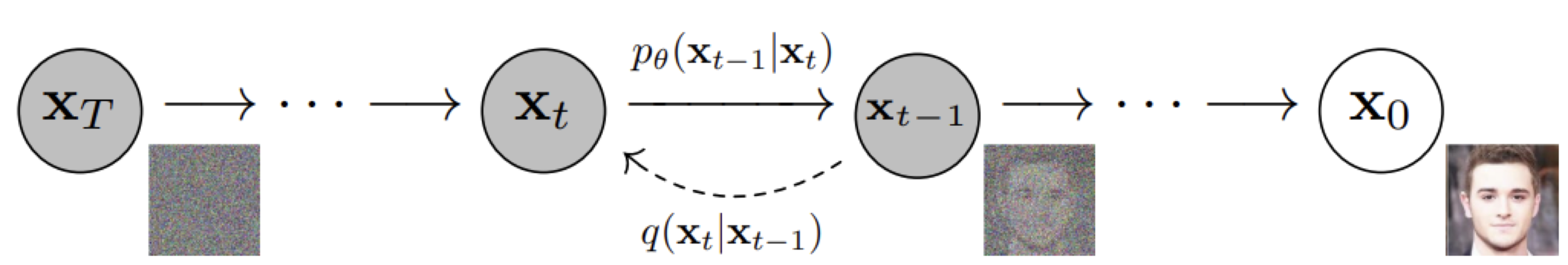}
    \caption{Denoising Diffusion Probabilistic Model Forward and Reverse Chain~\cite{ho2020denoising}.}
    \label{fig:ddpm}
\end{figure}

Diffusion models (DM) are inspired by non-equilibrium thermodynamics and stochastic differential equations~\cite{yang2023diffusion}. The DMs training process consists of two Markov Chains, a forward chain that adds noise at each step until it destroys the structure of the data to a simple noise. The reverse chain removes the noise at each step to reach the original data, as illustrated in Figure~\ref{fig:ddpm}. 
Unlike the parameters for the forward steps that are usually predefined as Gaussian Noise, leading the data distribution to be nearly an isotropic Gaussian distribution, the reverse process parameters are learned through a neural network to approximate the forward process by minimizing the KL-divergence between the joint distribution of the reverse process and the forward process~\cite{yang2023diffusion,ho2020denoising}. To condition the image generation on the prompt (text), the text is tokenized and converted to embedding, which are later transformed through a text transformer to be fed the noise predictor, which steers the reverse diffusion process toward the query text~\cite{nokey}. Although TTI systems can be called models, we opt to call them systems as their architecture involves multiple modules~\cite{friedrich2023fair}.
\section{Text-to-Image Representativity Fairness Evaluation Framework}
\label{sec:proposed_approach}
\begin{figure*}
    \centering
    \includegraphics[width=\linewidth]{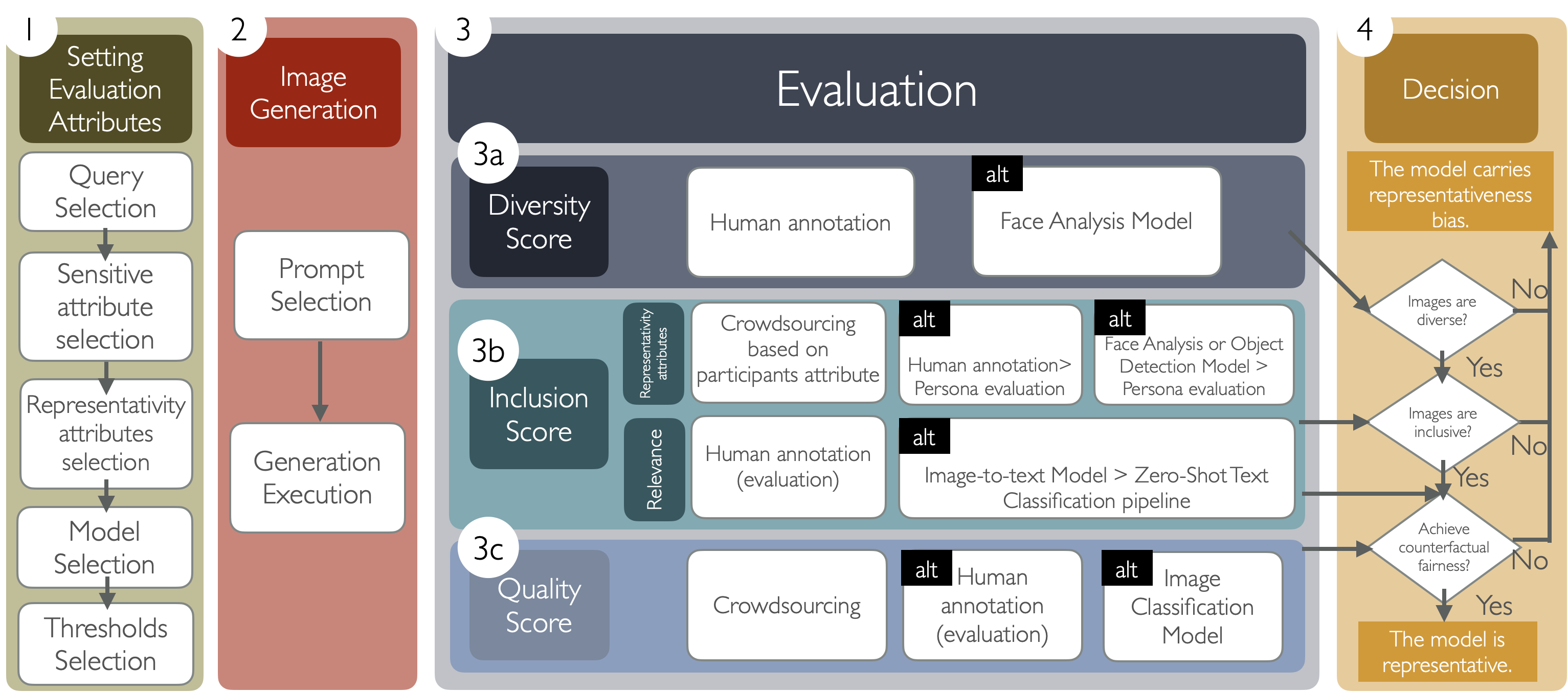}
    \caption{Text-to-Image Representativity Fairness Evaluation Framework}
    \label{fig:framework}
\end{figure*}

This section presents the proposed TTI Representativity Fairness Evaluation Framework, illustrated in Figure~\ref{fig:framework}, and discusses the approaches to quantifying the fairness metrics.


\subsection{Setting Evaluation Attributes} To start the evaluation, several evaluation subjects must be defined. 
    \begin{itemize}
        \item The query(s) $q$ is the pool of words within the studied context to be used in the prompt. They could be jobs (e.g., doctor, nurse) in the context of studying gender or race bias in jobs. They could be emotions (e.g., happy, angry, sad) in the context of combating stereotypes.
        \item The sensitive attribute under study $A$, defined by the regulatory entity. For example, $A$ can be race, age group, material status, or economic status. Also, the values of $A$ : $a_{n}$. For example, for $A=race$, the set ${A}=$\{Asian, Black, Caucasian, Indian, Latino, Middle Eastern\}
        \item The inclusion attributes $Inc\_F$ for testing for individual fairness and mitigating the issue of stereotypes. When studying race, the set $Inc\_F$ could consist of $Inc =$ \{age, gender, hair color, skin tone\}, and their respective attributes or ranges.
        \item The TTI system(s) to be studied.
        \item The thresholds for the diversity and inclusion scores, in addition to  the value of $\epsilon$ related to multi-class statistical parity.
    \end{itemize} 

\subsection{Image generation} A simple prompt should be used to avoid biases from other confounding factors. Two prompts have to be used, one without the sensitive attribute under study to test diversity and the other with the sensitive attribute value to test for inclusion. Multiple seeds and images per seed should be generated. The same seeds should be used when studying multi-class statistical parity to avoid the seed being a confounding factor. High computation and high RAM capabilities are needed for this process.

\subsection{Evaluation} Three measures are considered in this evaluation; diversity, inclusivity, and quality. Diversity and inclusion are adopted from~\cite{10.1145/3375627.3375832} and stem from considering the generation as a subset selection problem in the latent space. Quality here concerns having a photorealistic image or any other image qualities the user wishes to have.
To quantify the measures, two types of evaluation are present in the framework: Human-based and Model-based evaluations. In general, a human-based approach is resource expensive in terms of time and cost. It requires a representative sample of the population when using crowdsourcing or a prolonged duration of annotation when using a single expert annotator; in some cases, as will follow, an annotator from each member of the sensitive attribute is required for a fair evaluation. A human-based approach is resorted to for the sensitivity of the context or when no reliable and accurate model exists to identify the representativity attributes. On the other hand, a model-based approach  is assumed to be the least expensive, especially in the case of the availability of an open-source, highly-performing model. A human-in-the-loop approach helps mitigate any biases or inaccuracies of the model-based approach. 

\subsubsection{Diversity} The output images from $\hat{G}(q_{x})$ are considered with no mention of any of the instances of $A$. Two methods are considered to evaluate the generated images; a human-based approach and a model-based approach:
\begin{itemize}
    \item For the \textbf{human-based approach}, the sensitive attribute is annotated by a human or by reviewing the model's annotation from the model-based approach.
    \item For the \textbf{model-based approach},  a facial analysis model is used when $A$ is the race, age, or gender. 

\end{itemize}
As \cite{teo2021measuring} concluded, using one metric is insufficient for measuring the diversity of the output of generative models. Therefore, we employ both KL-Divergence as in \cite{MOWSHOWITZ2002141,tan2021improving} and total variational distance, which is more sensitive to minor differences. 

To bound KL-Divergence in the range [0, 1], the following transformation is performed
\begin{equation}
    tKL = exp(-D_{KL} \left( P \middle\| Q \right))
\end{equation}

For both metrics, $P(A)$ would be the distribution of the sensitive attribute, and $Q(A)$ is the uniform distribution. A score of zero indicates that the membership to the sensitive attribute does not affect its appearance as a generated output, hence a diverse output of the model. We complement the metrics results to measure the diversity.
\begin{equation}
    diversity\_  score_{(tKL)} = 1- tKL
\end{equation}

\begin{equation}
    diversity\_ score_{(TVD)} = 1- TVD
\end{equation}

\subsubsection{Inclusivity} This concept relates more to individual fairness and the output images from $\hat{G}(q_{x,a})$ are considered to measure the inclusivity.
The inclusivity score will be used for two goals; the first is to check if the model is inclusive for various members of the sensitive attributes under study with a certain threshold. The second is to check if multi-class statistical parity is achieved $\forall a \in A$. As in \cite{10.1145/3375627.3375832}, the inclusivity score is calculated as a function of the image's relevance to the query and statistical aggregation of the  inclusiveness score of each item in set attributes examined for representativeness $Inc\_F$. \par
To measure \emph{inclusivity of representativity attributes}, we propose a human-based approach, a model-based approach, and a human-in-the-loop approach. 
        \begin{itemize}
            \item For a \textbf{human-based approach}, the inclusivity of representativity attributes score can be calculated through crowdsourcing with ensuring representative participants $\forall a \in A$. The participant provides their $Inc\_F$ attributes and how they identify themselves with respect to $A$. Then, the participants are given multiple sets of images based on $A$ and are asked if any of the presented images meet one or more of their $Inc\_F$ characteristics. 
            \item For a \textbf{model-based approach}, a model would annotate the different $Inc\_F$ attributes of the generated images. This model could be an object detector or a face analysis model. Then personas with attributes drawn from a fair distribution, based on the context, are used to calculate the scores.
            \item A \textbf{human-in-the-loop-based approach} is to have a human annotate the $Inc\_F$ attributes from scratch or review the model's annotation, then use the same persona evaluation approach to calculate the inclusivity of representativity attributes scores. An aggregation method then combines the results of each attribute in $Inc\_F$. 
      \end{itemize}      
Calculating the representativeness score of each attribute in $Inc\_F$ differs from one attribute to another. It could be based on equality, a normalized difference of float or nominal attributes, or any other normalized measure between [0,1]. Nash inclusivity is used to aggregate the inclusion attributes scores, which is the geometric mean over the inclusion scores for each attribute $f$ in the set $Inc\_F$. 

\begin{equation}
\label{eq: nash}
  \text{attribute inclusivity score}(f,Inc\_F) = \sqrt[n]{ \prod_{f} score(Inc\_F_{f})}. 
\end{equation}
        
\par

To measure \emph{relevance}, we consider the following:
\begin{itemize}
    \item For a \textbf{human-based approach}, a human annotator can annotate the images based on defined criteria for relevance producing a score between $0$ and $1$.
    \item For a \textbf{model-based approach}, an image relevance pipeline is proposed consisting of an image-to-text model to generate a caption for the image followed by zero-shot multi-class classification. Based on the rank and/or confidence score with respect to other $x \in X$, the relevance score is calculated.
\end{itemize}

To aggregate the result of the relevance and inclusion attribute score, we propose following a utilitarian mechanism by averaging the outcome of both scores.

\begin{equation}
  \text{inclusivity score}(\hat{G}(q_{x,a})) = \frac{\text{representativity attributes score}(f,Inc\_F)_{a} +  \text{relevance score}_{a}}{2}
\end{equation}

\subsubsection{Quality} The output of $\hat{G}(q_{x,a})$ is considered. Two methods are considered to evaluate the generated images for quality:
\begin{itemize}
    \item For a \textbf{human-based approach}, crowdsourcing is proposed while ensuring representative participants $\forall a \in A$. The participant provides their $Inc\_F$ attributes and how they identify themselves with respect to $A$ and are given multiple sets of images based on $A$. The score is the ratio of the images they would use in a project to the total number of images in each set. This qualitative way of evaluating generated images is called preference judgment~\cite{borji2018pros} and has been used in~\cite{xiao2019generating, yi2018dualgan, zhang2017stackgan}.  Alternatively, a human annotator can be used with specific guidelines.
    \item  For a \textbf{model-based approach}, a classification model of the quality can be used per system.

\end{itemize}

\subsection{Decision} To draw a decision on whether bias exists in the TTI system, the diversity is considered first:
\begin{equation}
diversity\_ score = 1 - \epsilon \Rightarrow \texttt{Diverse Model},    
\end{equation}
where $\epsilon$ is a threshold set by the regulatory entity or if absent the reviewer. We start with diversity as it is a necessary condition to study inclusion. A non-diverse model will be marked as a model containing representativity bias.\par
Following, inclusivity is considered per $a$ based on the aggregate of the inclusivity of representativity attributes and relevance scores:

\begin{equation}
\forall a \in A , \text{inclusivity score}(\hat{G}(q_{x,a})) > \epsilon \Rightarrow \texttt{Inclusive Model},      
\end{equation}

If $\hat{G}$ passes as an inclusive model, then the model is examined for multi-class statistical parity for both inclusivity and quality:

\begin{equation}
\forall a\in A |P(\text{inclusivity score}(\hat{G}(q)|A=a) - \mathop{\mathbb{E}}_{a\in A}(\text{inclusivity score}\hat{G}(q)|A=a)| \leq \epsilon,
\end{equation} and 

\begin{equation}
\forall a\in A |P(\text{quality score}(\hat{G}(q)|A=a) - \mathop{\mathbb{E}}_{a\in A}(\text{quality score}\hat{G}(q)|A=a)| \leq \epsilon.
\end{equation} 

 If the inequalities are satisfied, then the generative model holds multi-class statistical parity for the sensitive attribute $A$ and Representativity Fairness for the context under study.
\section{Experimental Evaluation}
\label{sec:exp}
This section presents the experimental design and results of evaluating a TTI system using the TTI Representativity Fairness Evaluation Framework.
\subsection{Setting Evaluation Attributes}
We start with the study design by specifying the four elements in the context of evaluating a TTI system for auditing racial bias in images for occupations:
\begin{enumerate}
    \item Query: $q$ focus on occupations and varies between high-paying jobs and low-paying jobs. It also includes male-majority and female-majority jobs. It includes the following set: $q$ = \{CEO, salesperson, scientist, janitor, engineer,
     Electrician, Plumber, Carpenter, Driver, 
     Secretary, Receptionist, Maid, Nurse, 
     Doctor, Pilot,  Lawyer, Software developer, 
     Farmer, Baker, Hairstyles, Desk clerk, cashier'\}. T
     \item The sensitive attribute under study is race. $A=$ \{Asian, Black, Caucasian, Indian, Latino, Middle Eastern\}
     \item Representativity attributes:  $Inc\_F =$ \{ gender: \{women,man\}, age: [15,65] \} 
     \item TTI system: Stable Diffusion $_{base}$ \cite{Rombach_2022_CVPR}, version v2.1\footnote{https://huggingface.co/stabilityai/stable-diffusion-2-1-base}, using the default parameters and with a guidance scale of $7$.
     \item No reported regulations are reported for the domain. Therefore, we set some lenient thresholds of $0.70$ for the diversity score, $\epsilon = 0.15$ for multi-class statistical parity, and $0.55$ for the overall inclusion score.   
\end{enumerate}

\subsection{Image Generation} Images were generated using the Stable Diffusion Pipeline on Google Colab Pro utilizing an NVIDIA 100-SXM GPU with 25 GB RAM. Five images were generated per (query, seed). Two simple prompts structures were used: To measure diversity: 'A $worker_{i}$.', and following \cite{friedrich2023fair} 'A photo face of a $worker_{i}$.'. For inclusivity and quality, the prompts are conditioned on $A$ as in 'A $a$ $worker_{i}$.' and 'A photo face of a $a$ $worker_{i}$.'. $19$ random seeds were considered to measure diversity, while $3$ were used in prompts conditioned on $a$, $\forall a \in A$. The total generated images is $4180$ for prompts of $q_{x}$ and $1320$ for prompts of $q_{x, a}$. 

\begin{figure}
     \centering
     \begin{subfigure}[b]{0.25\textwidth}
         \centering
         \includegraphics[width=\textwidth]{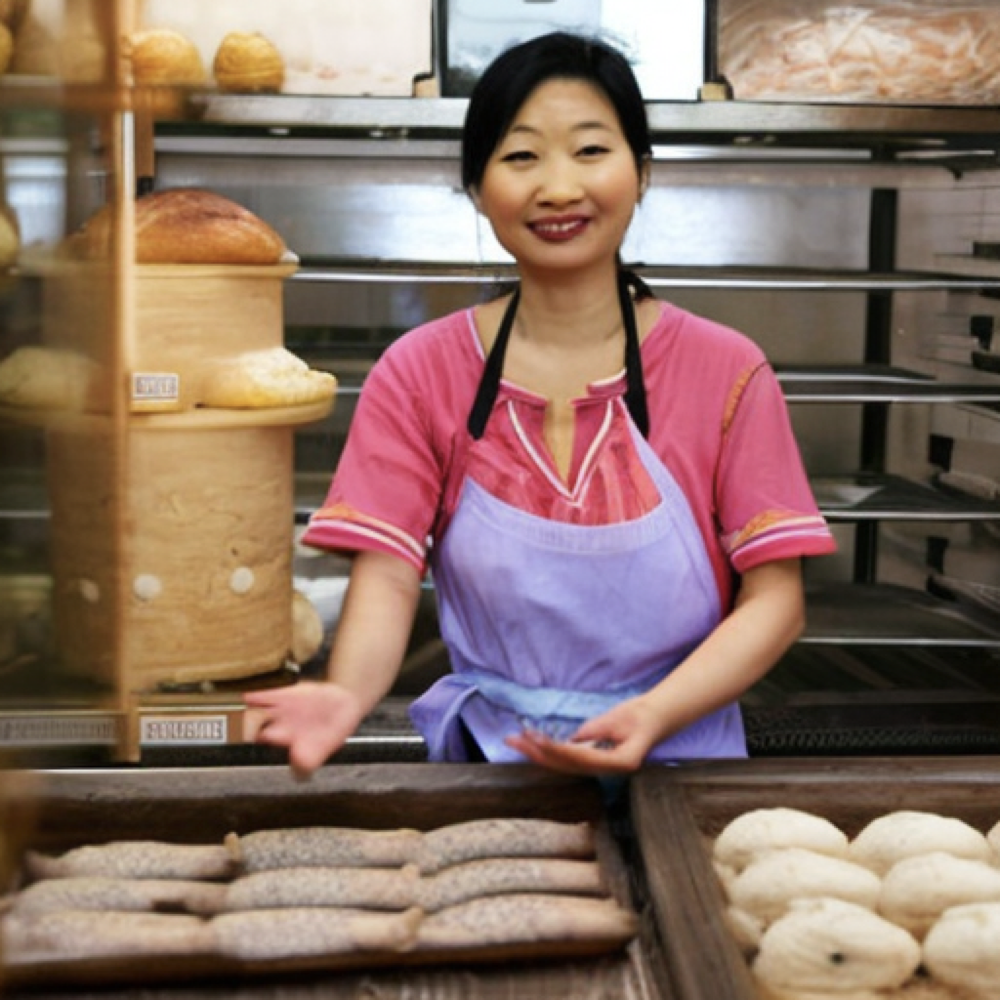}
         \caption{Asian Baker}
     \end{subfigure}
     \begin{subfigure}[b]{0.25\textwidth}
         \centering
         \includegraphics[width=\textwidth]{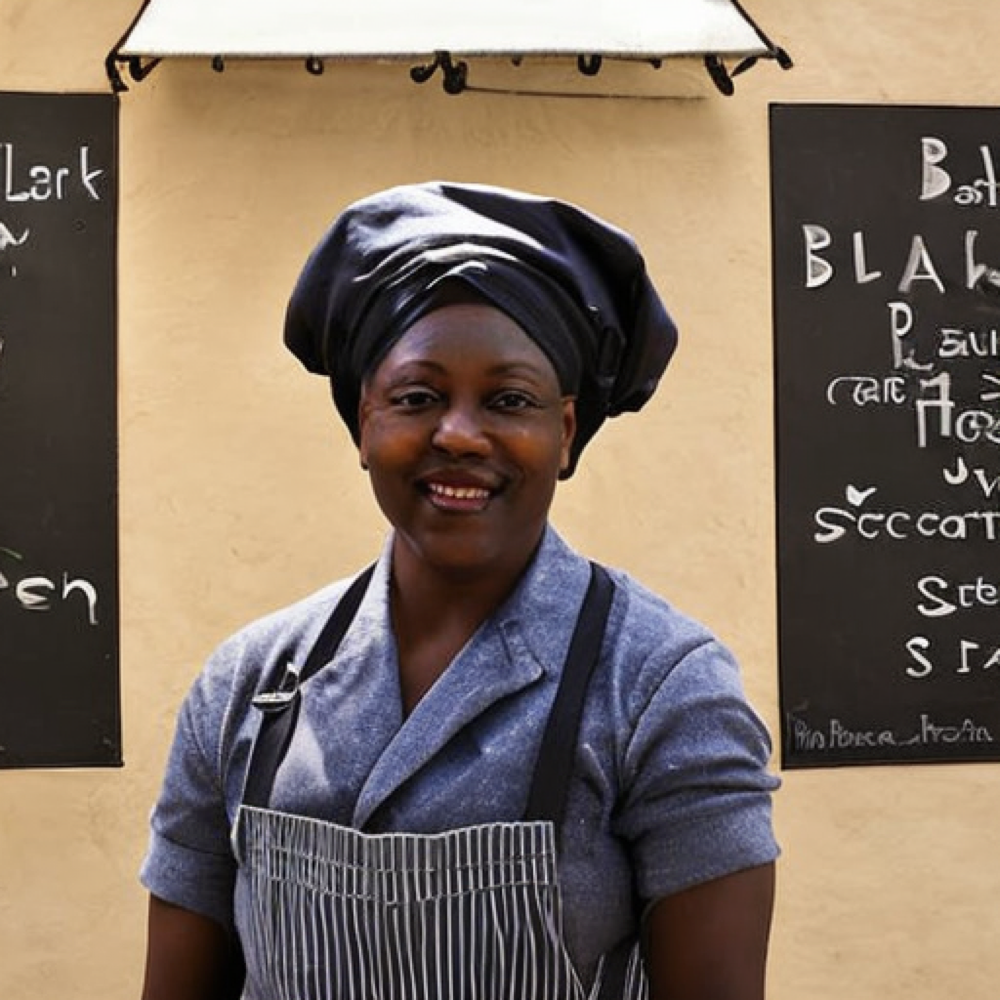}
         \caption{Black Baker}
     \end{subfigure}
          \begin{subfigure}[b]{0.25\textwidth}
         \centering
         \includegraphics[width=\textwidth]{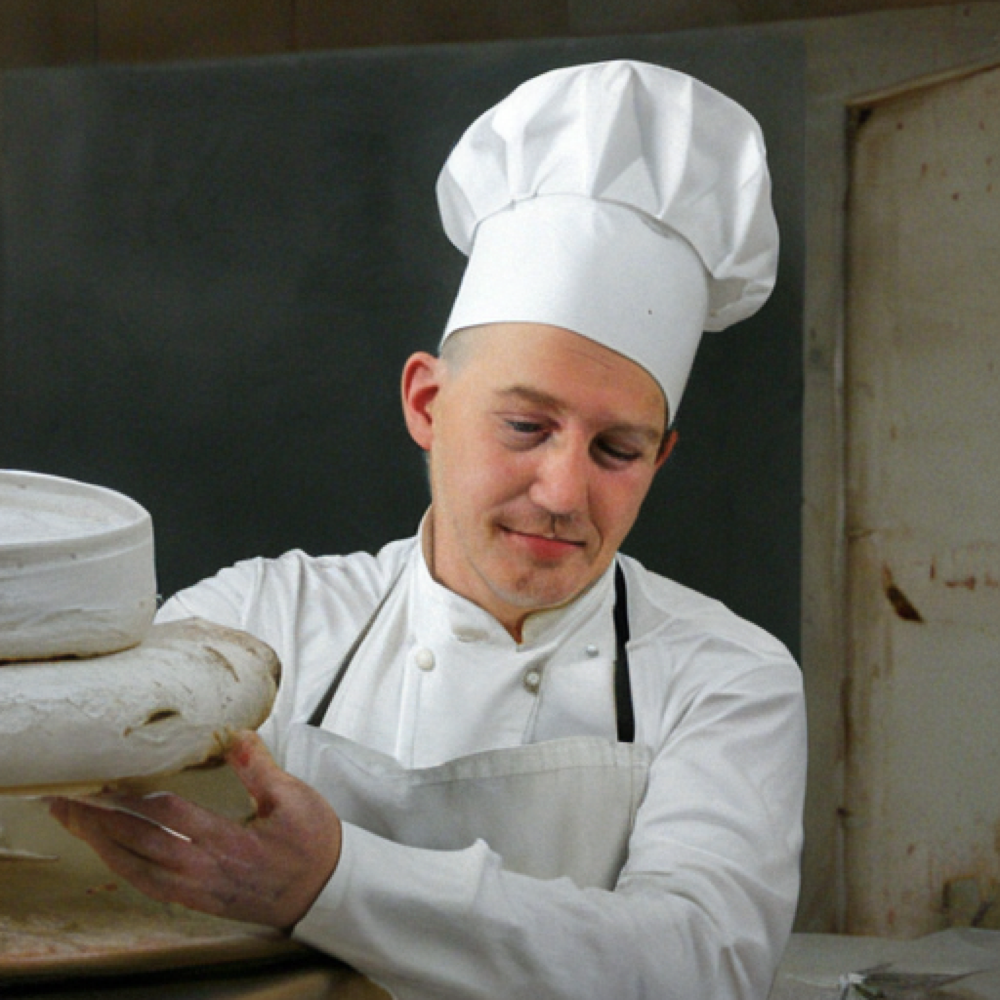}
         \caption{Caucasian Baker}
     \end{subfigure}
     \begin{subfigure}[b]{0.25\textwidth}
         \centering
         \includegraphics[width=\textwidth]{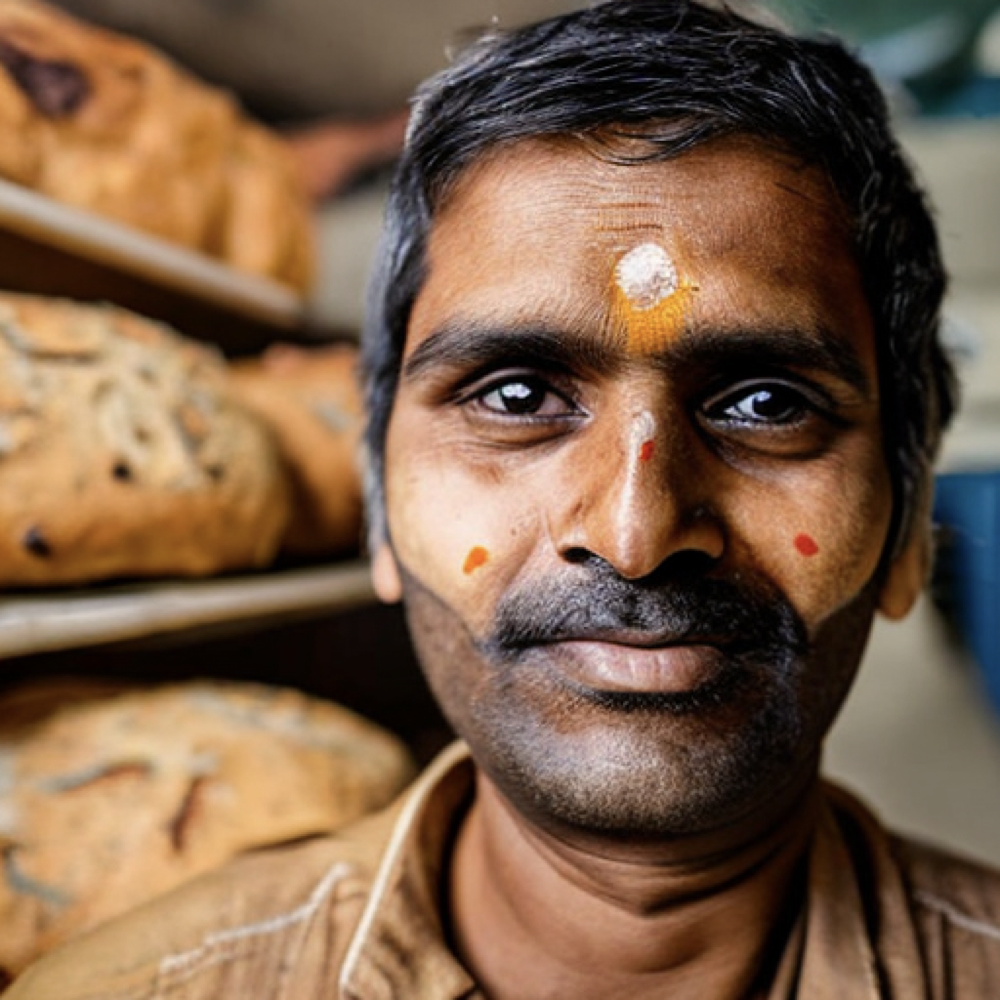}
         \caption{Indian Baker}
     \end{subfigure}
     \begin{subfigure}[b]{0.25\textwidth}
         \centering
         \includegraphics[width=\textwidth]{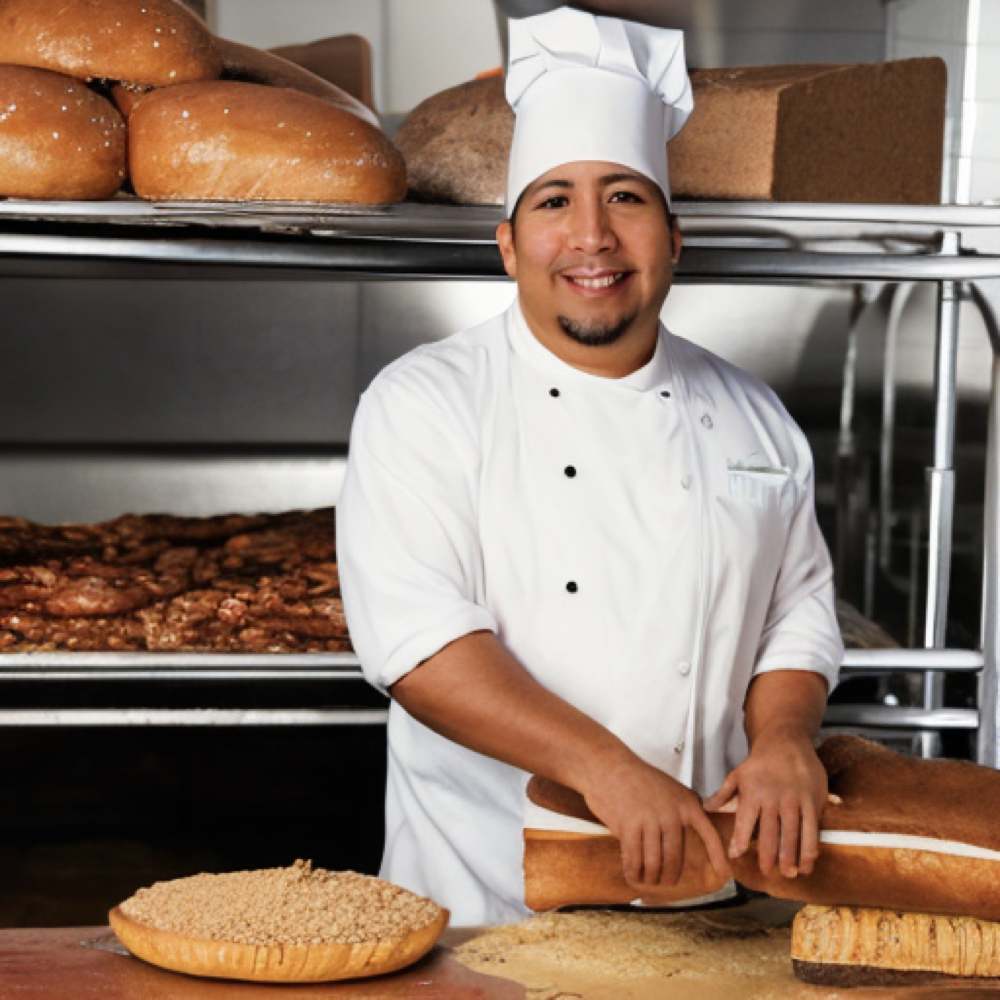}
         \caption{Latino Baker}
     \end{subfigure}
     \begin{subfigure}[b]{0.25\textwidth}
         \centering
         \includegraphics[width=\textwidth]{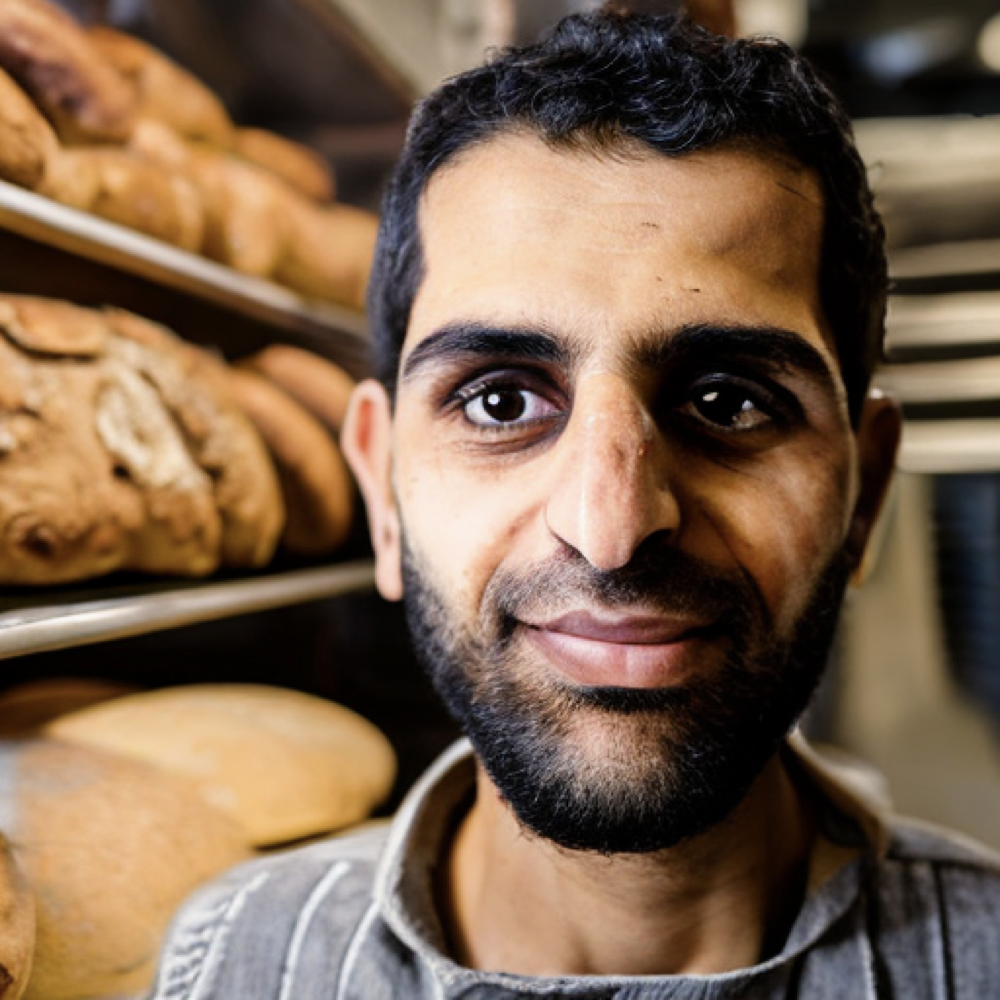}
         \caption{Middle Eastern Baker}
     \end{subfigure}

        \caption{Baker of different races.}
        \label{fig:baker}
\end{figure}

To contain the scope of the study, we only annotated and performed further analysis on the occupations \{doctor, cashier, baker, secretary, software engineer, lawyer\}. This subset of jobs still maintains low/high paying jobs and male/female majority jobs.  Also, as including the phrase 'photo face of a $worker_{i}$ ' yields clearer photos, only the second and fourth prompts were considered. Hence, $570$ images were considered to study the diversity and $540$ to study inclusion and inclusion multi-class statistical parity and quality, totaling $1110$ images. Figure~\ref{fig:baker} presents a sample of generated images for bakers. All images are available upon request.

\subsection{Evaluation}
\subsubsection{Diversity} We selected the Uniform distribution as our fair distribution $Q(A)$. As $|A| = 6$, $\forall a \in A Pr(A) = \frac{1}{6}$. Also, we started with the model-based evaluation, followed by the human-based evaluation.\par

\begin{figure}
     \centering
     \begin{subfigure}[b]{0.7\textwidth}
         \centering
         \includegraphics[width=\textwidth]{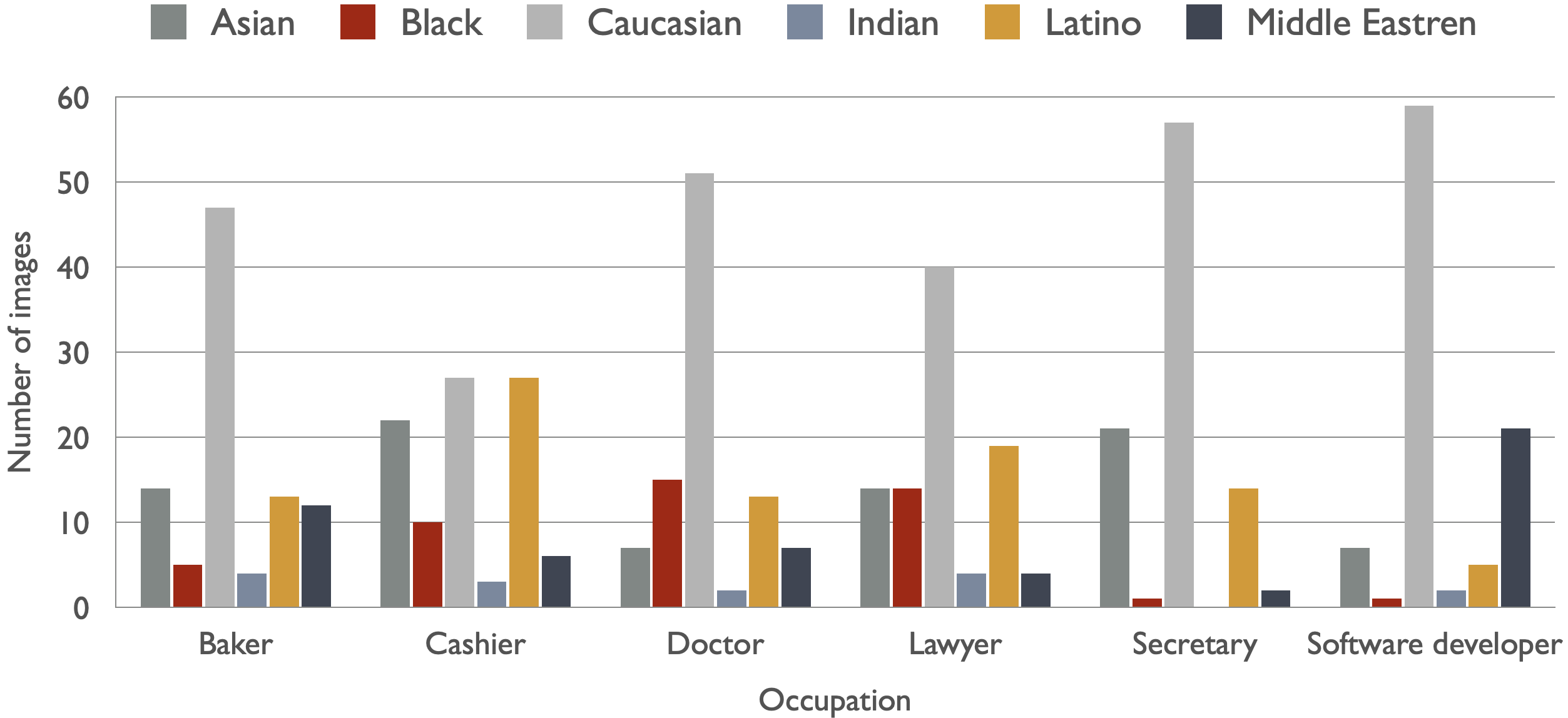}
         \caption{Model-based evaluation}
     \end{subfigure}
     \begin{subfigure}[b]{0.7\textwidth}
         \centering
         \includegraphics[width=\textwidth]{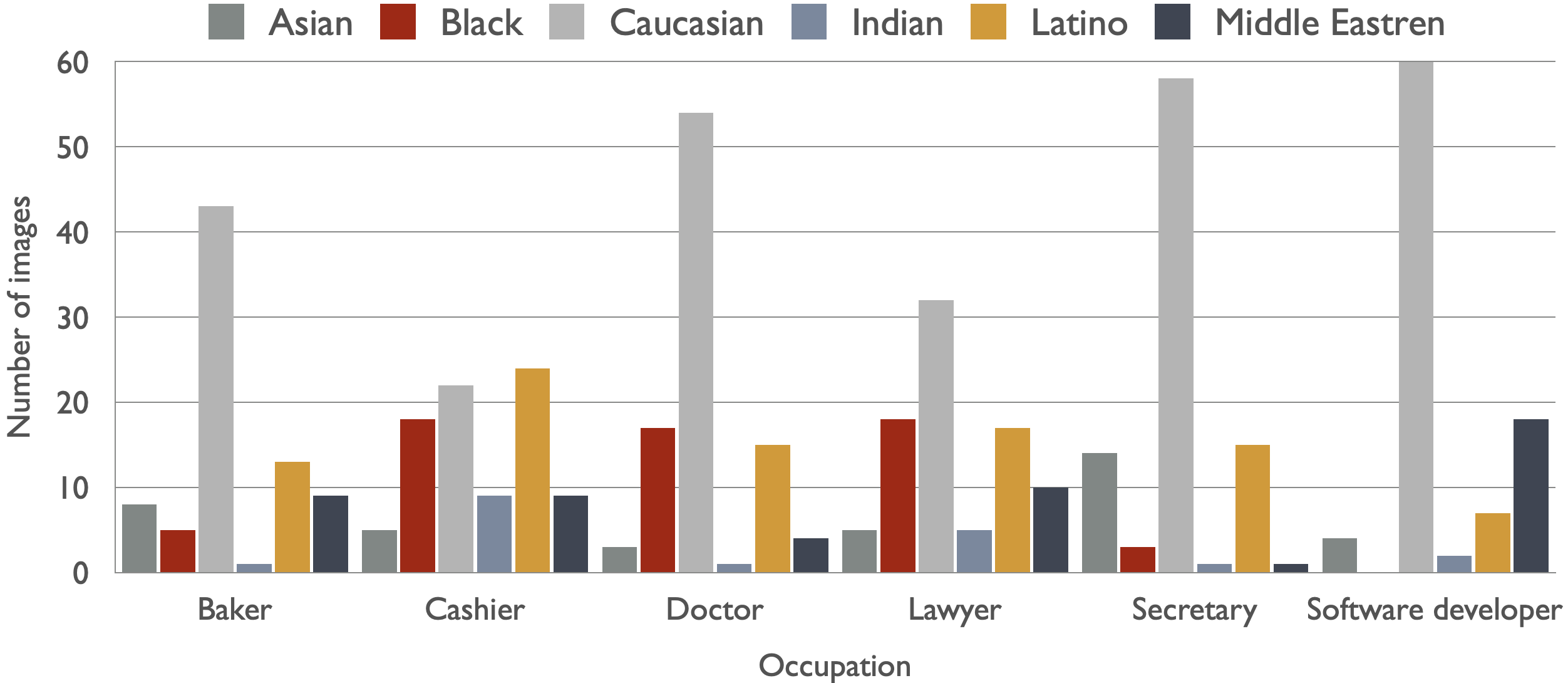}
         \caption{Human-based evaluation}
     \end{subfigure}
        \caption{Distribution of race across generated images.}
        \label{fig:diversity}
\end{figure}

\begin{table}[]
\caption{Diversity score based on Human and Model Evaluation and using KL-Divergence and TVD.}
\label{tab:diversity}
\begin{tabular}{@{}lllll@{}}
\toprule
                     & \multicolumn{2}{l}{Diversity Score$_{KL-Divergence}$} & \multicolumn{2}{l}{Diversity Score$_{TVD}$} \\ \midrule
                     & Model                      & Human                     & Model                 & Human                \\ \midrule
Baker                & 0.712                      & 0.635                     & 0.672                 & 0.622                \\
Cashier              & 0.805                      & 0.879                     & 0.700                 & 0.764                \\
Doctor               & 0.651                      & 0.560                     & 0.630                 & 0.578                \\
Lawyer               & 0.760                      & 0.817                     & 0.712                 & 0.730                \\
Secretary            & 0.477                      & 0.492                     & 0.512                 & 0.536                \\
Software Developer   & 0.504                      & 0.453                     & 0.491                 & 0.473                \\ \midrule
Overall   & 0.65                      & 0.63                     & 0.67                 & 0.65                \\ \midrule

Spearman Correlation & \multicolumn{4}{c}{0.942}                                                                             \\ \bottomrule
\end{tabular}
\end{table}



\emph{Model-based approach.}
DeepFace~\cite{serengil2021lightface}, an open-source library for face recognition and analysis including age, race, gender, and emotions, was used to label the generated images for race~\footnote{https://github.com/serengil/deepface} when considering Model Annotation. The results in Figure~\ref{fig:diversity} show the distribution of the generated image. The overall diversity score based on KL-divergence is $0.65$. Table~\ref{tab:diversity} presents the breakdown of the diversity score based on the two metrics across occupations. As apparent from Figure~\ref{fig:diversity} and reflected in Table~\ref{tab:diversity}, the model is biased towards generating faces with Caucasian features. This tendency varies by job and is least apparent in a cashier and lawyer job.
 \par
\emph{Human-based approach.}
Human annotation was done by reviewing the model annotation and correcting any faulty annotation. Out of the $570$ images, $38$ were not assigned to a race due to the lack of face presence or poor quality, $18$ of them were mislabeled by the model to be white, and the rest were almost evenly distributed across the other races. The results in Figure~\ref{fig:diversity} show the distribution of the generated image. Table~\ref{tab:diversity} presents the diversity score based on the two metrics and occupation. The overall diversity score based on KL-divergence is $0.63$. As with the model-based approach, the model is biased toward generating faces with Caucasian features, especially for software developers and secretaries. \par
The model-based approach highly correlates with the human-based approach at $0.94$ Spearman Correlation, indicating that the model-based approach can substitute for the human-based approach in this context.



\subsubsection{Inclusion}
Inclusiveness is evaluated in two parts: relevance and inclusion of representativity attributes. For the first, we performed human-based annotation via a single annotator and model-based annotation through an Image Relevance pipeline. We have three options for the latter: in the first and second, we annotate the images for the representativity attributes through a human annotator or a model. For both options, we evaluate the inclusiveness through personas. The third option is crowdsourcing. 

\textbf{Relevance}
\begin{figure}
     \centering
     \begin{subfigure}[b]{0.45\textwidth}
         \centering
         \includegraphics[width=\textwidth]{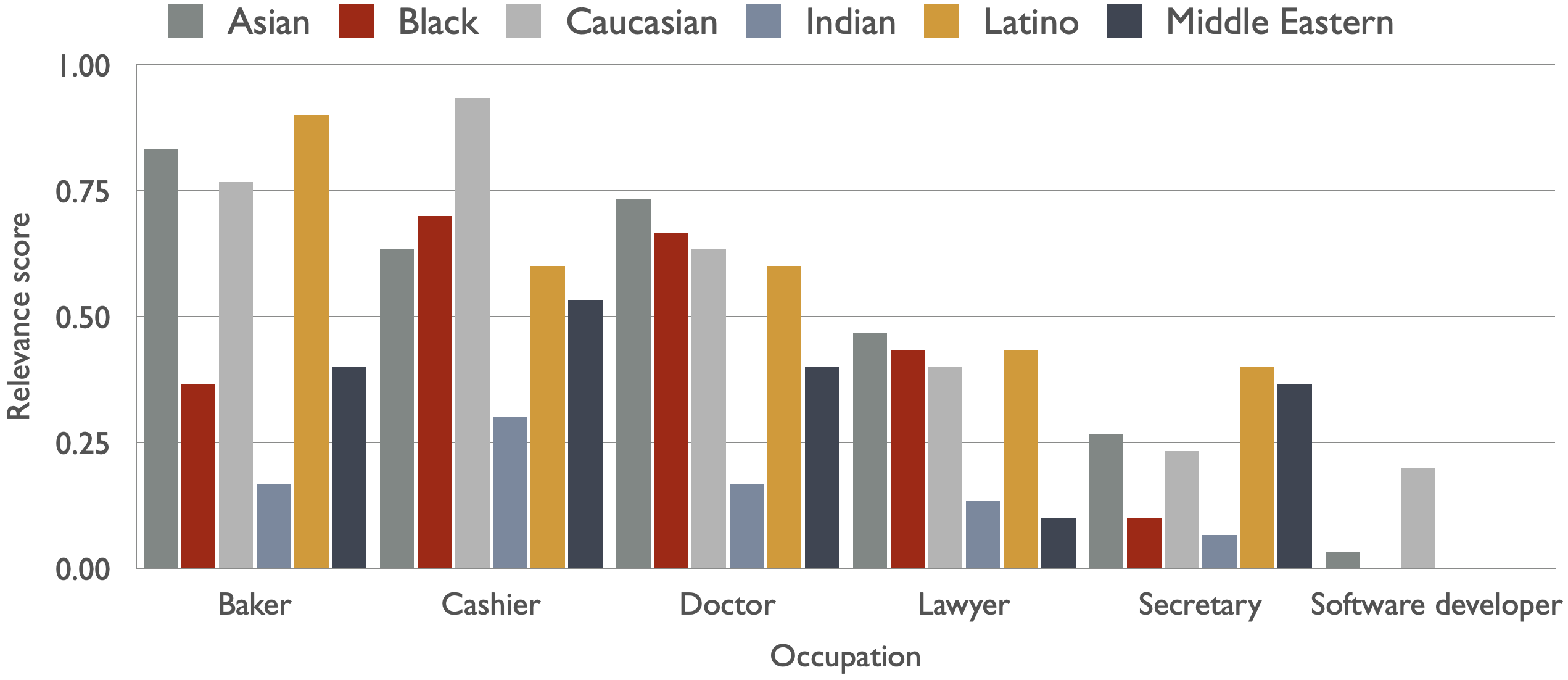}
         \caption{Single-human evaluation}
     \end{subfigure}
               \begin{subfigure}[b]{0.45\textwidth}
         \centering
         \includegraphics[width=\textwidth]{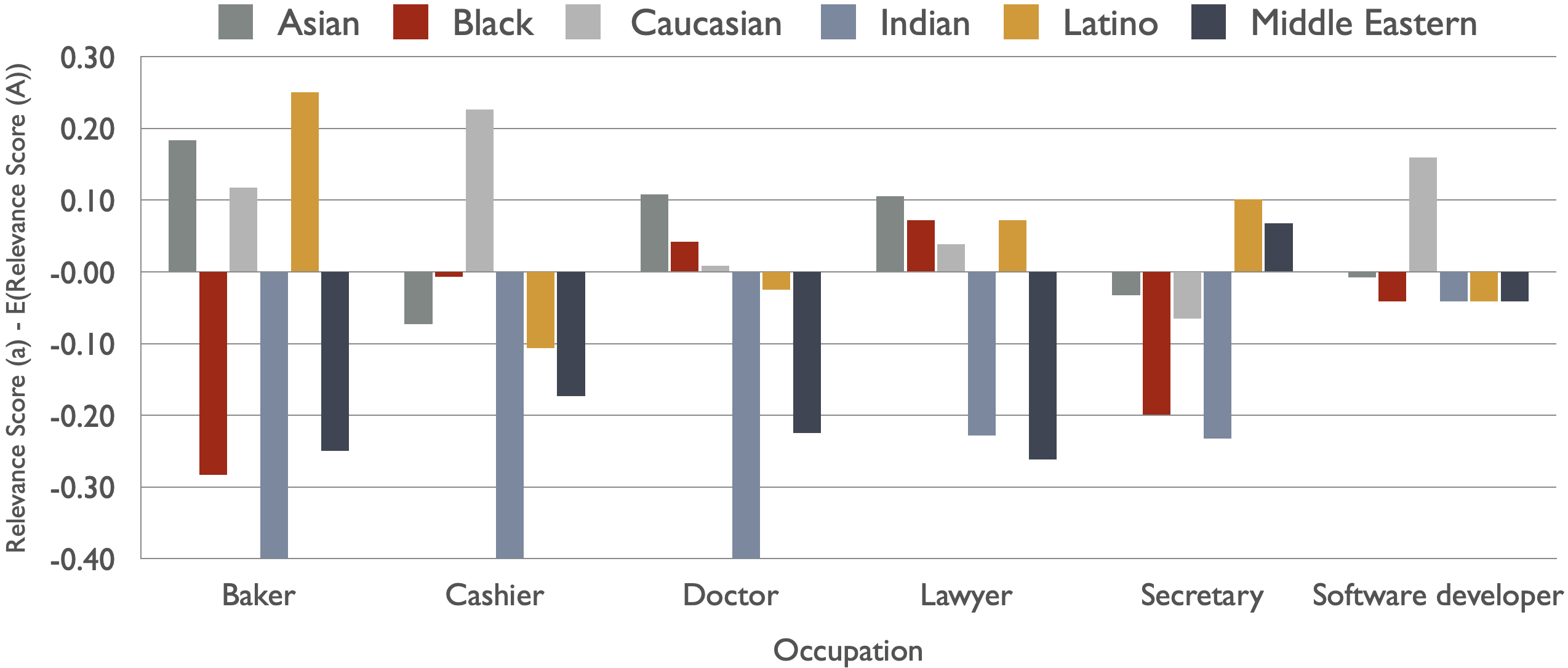}
         \caption{Score difference with human-based evaluation}
     \end{subfigure}

     \begin{subfigure}[b]{0.45\textwidth}
         \centering
         \includegraphics[width=\textwidth]{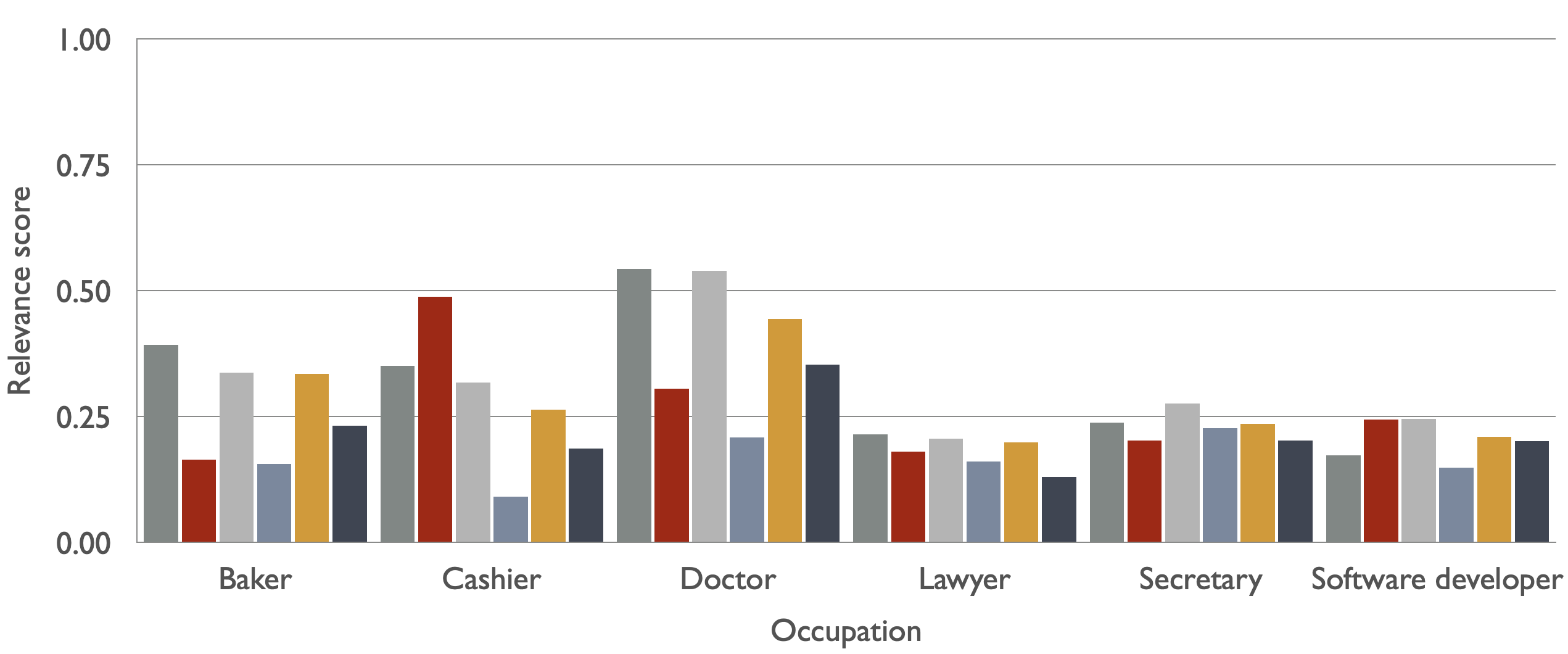}
         \caption{Image Relevance pipeline}
     \end{subfigure}
     \begin{subfigure}[b]{0.45\textwidth}
         \centering
         \includegraphics[width=\textwidth]{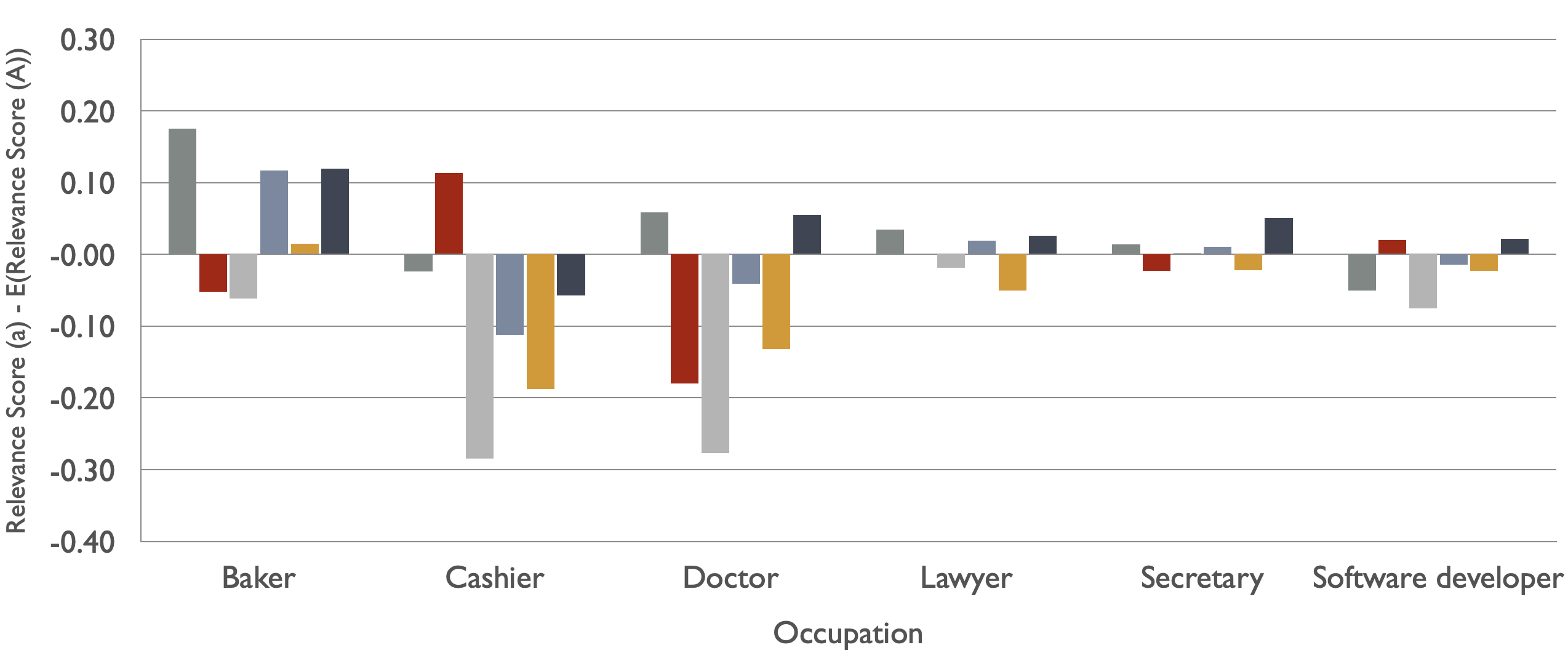}
         \caption{Score difference with relevance pipeline evaluation}
     \end{subfigure}

        \caption{Relevance scores across generated images.}
        \label{fig:relevance}
\end{figure}

\emph{Human-based approach.} A single annotator from the Middle East annotated the images for relevance with scores from zero to one. Zero implies there is no relevance at all, half means the relevance is in the attire only, and one implies relevance in the attire and at least one relevant object. The results of the relevance evaluation are in Figure~\ref{fig:relevance}. As we can notice from the scores, regardless of race, the relevance of the output varies significantly by job. Jobs having relevance scores more than average are cashier, baker, and doctor. Regarding race, Blacks, Middle Easterners, and Indians are more likely to have irrelevant images.

\emph{Model-based approach.} An Image Relevance pipeline is used consisting of two models; an image-to-text model that provides a caption to the image and a zero-shot text classification model. For the image-to-text classification model we used \emph{GenerativeImage2Text
 (GIT) model}~\cite{wang2022git}. The selection was made after experimenting with multiple models, including BLIP and VIT-GPT2 \footnote{https://huggingface.co/spaces/nielsr/comparing-captioning-models}, in which \emph{GIT} gave the most detailed description in different trials. The caption generated is neutralized by replacing a woman or man with a person. The neutralized caption is fed to an \emph{NLI-based Zero Shot Text Classification}~\cite{yin2019benchmarking} \footnote{https://huggingface.co/facebook/bart-large-mnli}, with a multi-class option set to true. The classes provided were the set occupations under study. The confidence score of the class prediction is used for relevance after mapping to ranks considering a confidence score above $0.3$ as $1$ (relevant), less than $0.2$ as $0$ (irrelevant), and anything in between as $0.5$ (medium relevance). The results of the relevance are in Figure~\ref{fig:relevance}. The score differences between the occupations and overall relevance are reduced as illustrated. Jobs having relevance scores more than average now are cashiers and doctors only. When considering race, the differences are minimal, with Middle Eastern and Indians more likely to have irrelevant images, as illustrated in Figure~\ref{fig:relevance}. 
 
The image relevance pipeline correlates with the human-based approach at 0.70 Spearman Correlation, indicating that the image relevance pipeline can substitute for the human-based approach, especially in a non-critical context. 


\par
\textbf{Representativity Attributes.} Evaluating the inclusion of representativity attributes ensures that the model avoids stereotyping and produces representative images of the community under study in our experiment for each race. We consider two representativity attributes; age and gender. Three methods are proposed to evaluate representativity attributes. The first and second include testing the inclusion of representativity attributes against a persona with the difference in the annotation method. The third alternative is crowdsourcing.


\begin{figure}
     \centering
          \begin{subfigure}[b]{0.45\textwidth}
         \centering
         \includegraphics[width=\textwidth]{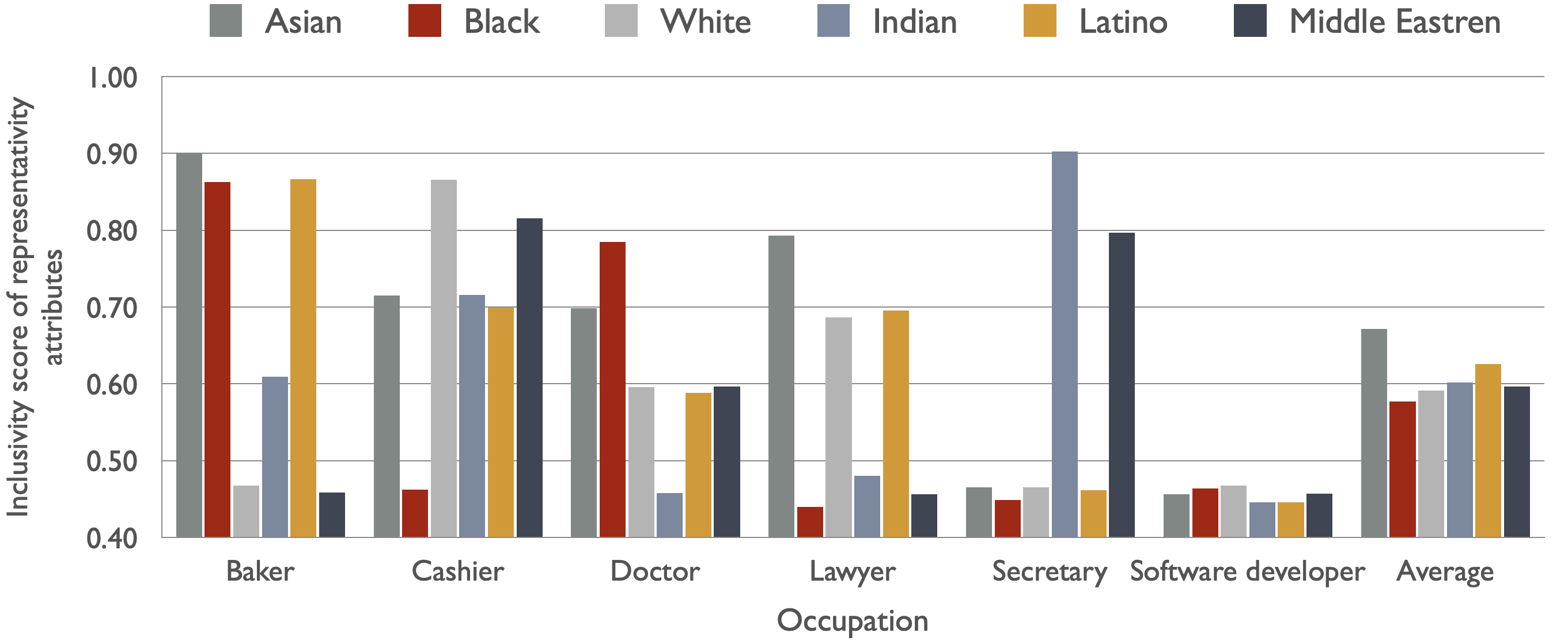}
         \caption{Model annotation with persona}
     \end{subfigure}
          \begin{subfigure}[b]{0.45\linewidth}
         \centering
         \includegraphics[width=\textwidth]{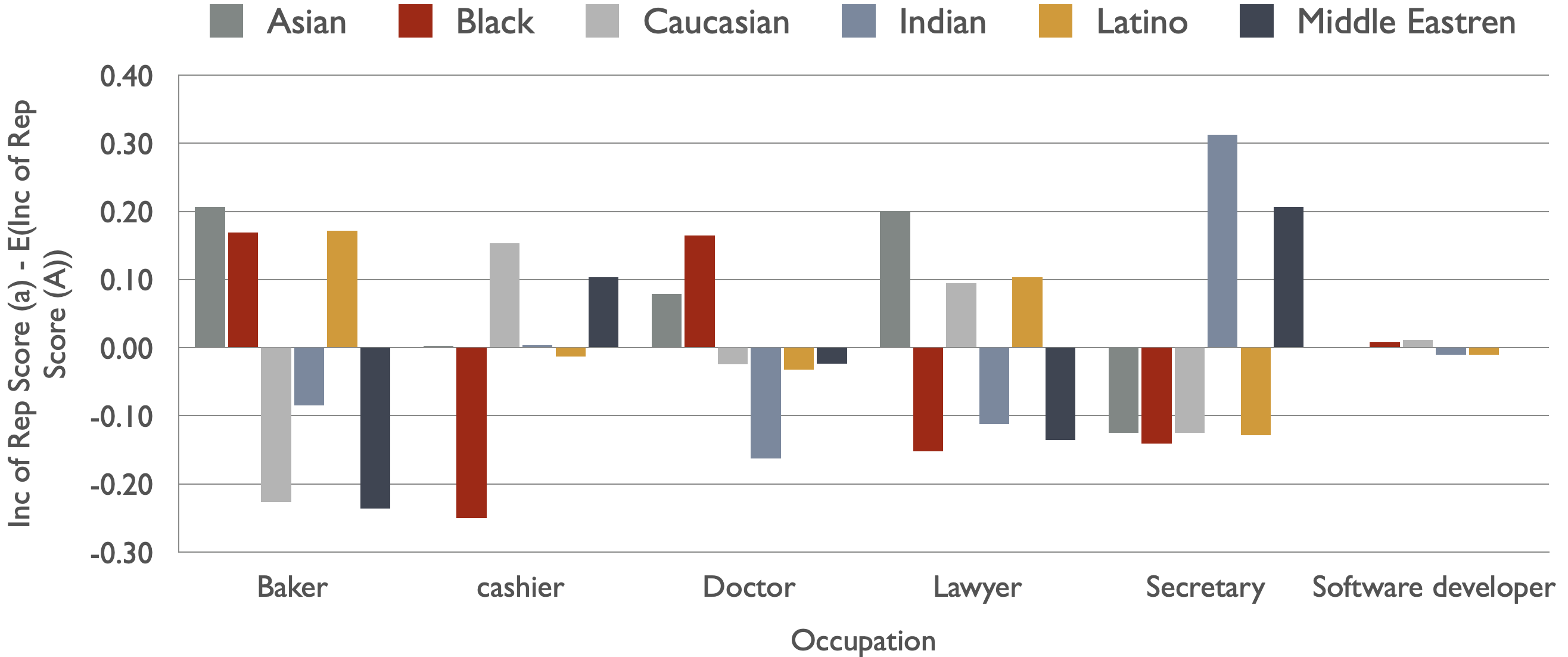}
         \caption{Score difference of the Model annotation with Persona}
     \end{subfigure}

     \begin{subfigure}[b]{0.45\textwidth}
         \centering
         \includegraphics[width=\textwidth]{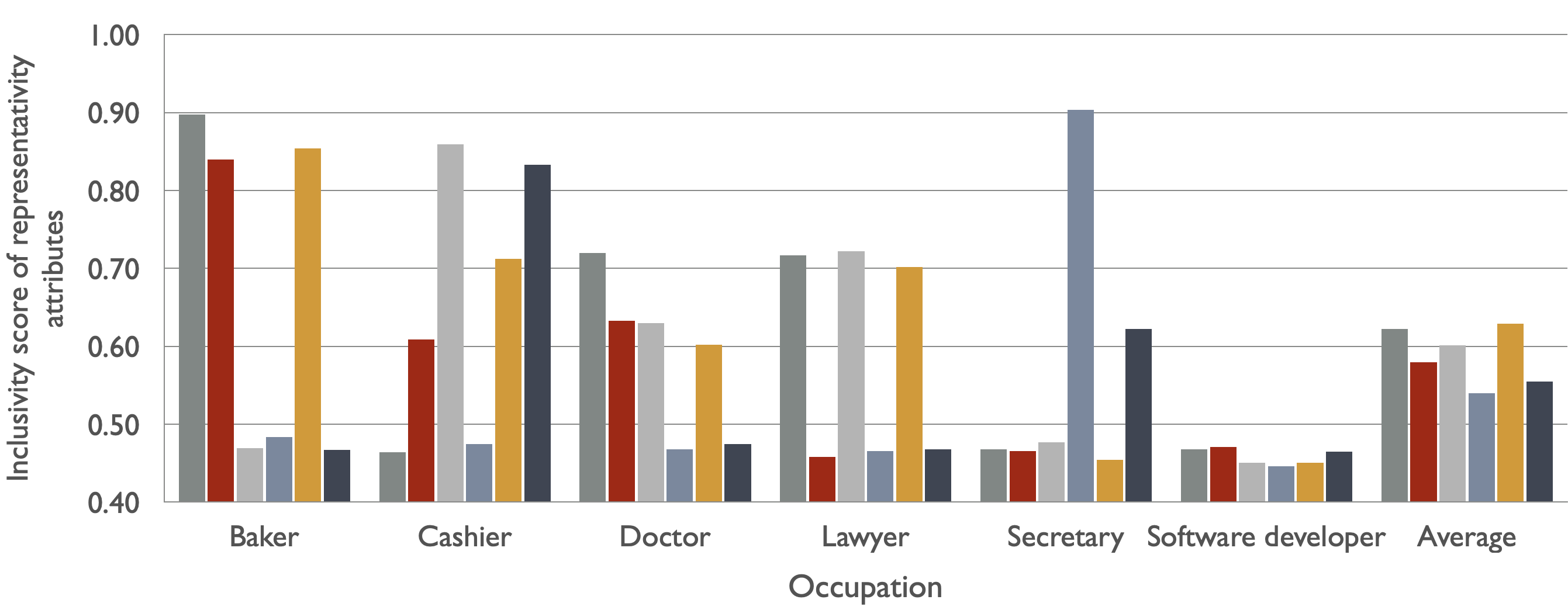}
         \caption{Human annotation with persona}
     \end{subfigure}
     \begin{subfigure}[b]{0.45\linewidth}
         \centering
         \includegraphics[width=\textwidth]{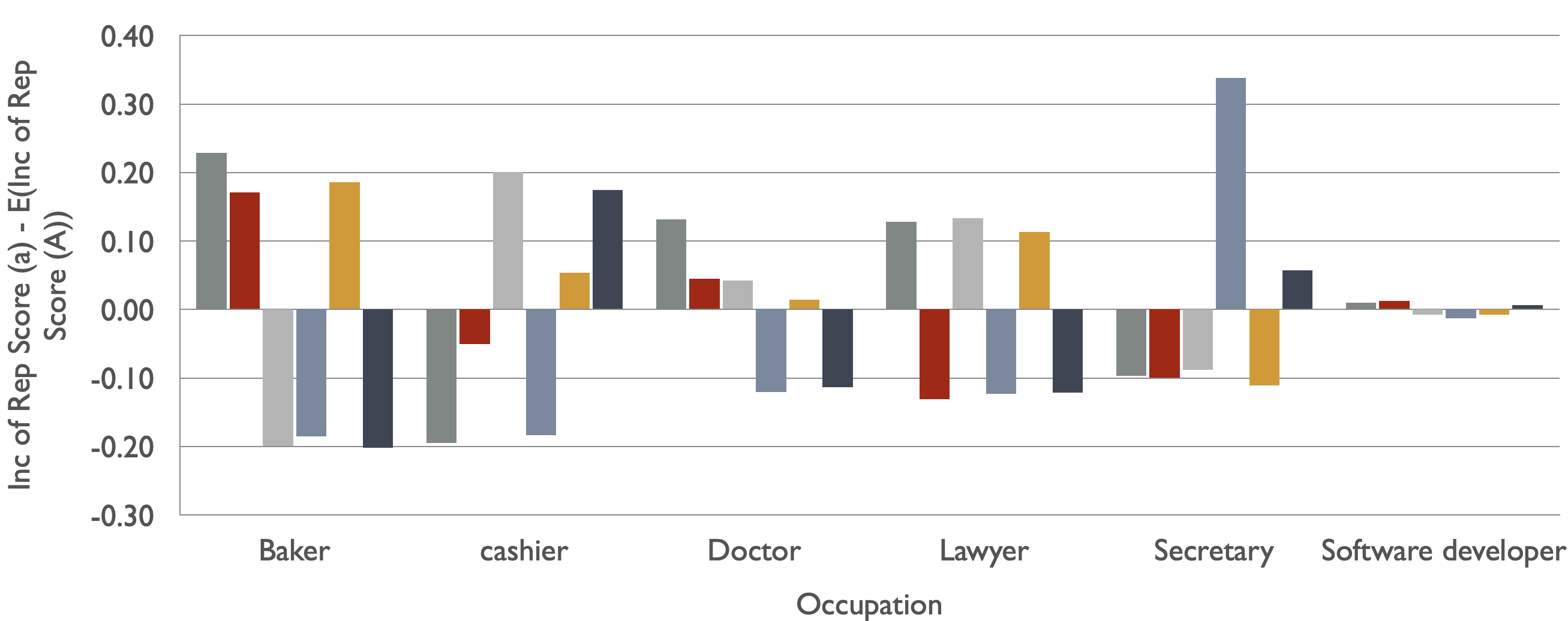}
         \caption{Score difference of the Human annotation with Persona}
     \end{subfigure}

        \caption{Inclusion of representativity attributes scores.}
        \label{fig:rep-scores}
\end{figure}

\emph{Model-based approach.} Persona with Model annotation via DeepFace~\cite{serengil2021lightface} was used to label the generated images for age~\footnote{https://github.com/serengil/deepface}. As for gender, the captions generated by \emph{GenerativeImage2Text
 (GIT) model}~\cite{wang2022git} were used. Whenever GIT did not provide gender markers in the caption generated (e.g., woman, man, boy, girl), DeepFace~\cite{serengil2021lightface} was used. $5000$ personas were generated with age and gender drawn from a uniform distribution. For age, the range of the uniform distribution was [15,65]. As for gender, we consider the binary gender of \{male, female\}. Five images are selected randomly to calculate their Nash score for each Persona  for each race and occupation. The highest Nash score per image, Equation \ref{eq: nash}, was considered the inclusion of representativity attributes score for the image set. For gender attribute, $1$ is awarded if the images match the gender of the persona, $0$ otherwise. For age, it is calculated as in Equation \ref{eq:sc_age}.
\begin{equation}
\label{eq:sc_age}
    score_{age} = 1 - \frac{|persona\_age - generated\_person\_estimated\_age|}{persona\_age\_range}
\end{equation}

The results in Figure~\ref{fig:rep-scores} show the distribution of the inclusion of representativity attributes scores across races and jobs. The software developer job receives the lowest average inclusion of representativity attributes scores across races; no images of women software developers were generated. Blacks and Caucasian ethnicities are on the lower end of the inclusion of representativity attributes scores, as shown in Figure~\ref{fig:rep-scores}, averaging $0.57$ and $0.59$ across occupations. Below the average also were Middle Eastern and Indians, averaging $0.60$ for both races.

\emph{Human-in-the-loop approach.} Persona with Human annotation was done by reviewing the model annotation and correcting any faulty annotation. Age was corrected only if more than a $5$ years difference is apparent. Persona generation and inclusion of representativity attributes scores calculation was performed similarly to the model-based annotation.\par

The results in Figure~\ref{fig:rep-scores} show the distribution of the inclusion of representativity attributes scores across races and jobs. Similarly to model-based annotation, the Software developer job receives the lowest average inclusion of representativity attributes scores across races. Indian and Middle Eastern ethnicities now have the lowest inclusion of representativity attributes scores, as shown in Figure~\ref{fig:rep-scores}, averaging $0.54$ and $0.55$ across occupations, respectively.

A Pearson correlation of $0.88$ and a Spearman correlation of 0f $0.82$ was achieved regarding the inclusion of representativity attributes scores for the human and model-based annotation with persona evaluation, indicating the ability to substitute the model-based approach for the human-based approach in this context.

\begin{figure}
 \centering
 \includegraphics[width=0.6\textwidth]{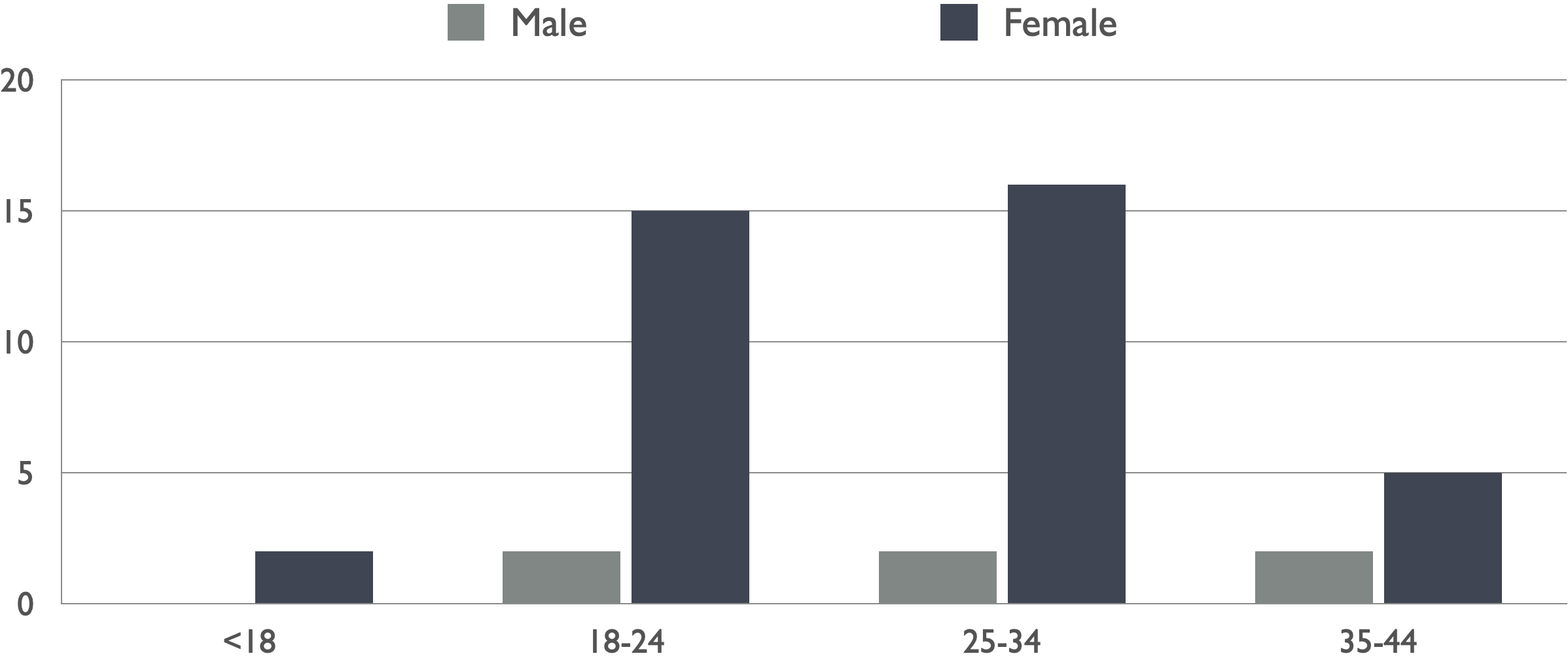}
\caption{Distribution of the Middle Eastern participants.}
\label{fig:questionnaire}
\end{figure}

\emph{human-based approach} Using crowdsourcing, a questionnaire was disseminated through social media platforms. We displayed three sets of images per occupation, each containing five images for the participant according to the race they identify with. Out of $51$ respondents, the majority, $44$, were middle eastern, three were Black, another three were South Asian, and one was Caucasian. We will only consider the Middle Eastern participants of the questionnaire as the responses from other races were limited. The distribution of the Middle Eastern participants is in Figure~\ref{fig:questionnaire}. To simplify the process, the participants were asked to indicate if the most inclusive image is inclusive towards either their age, gender, both, or none. A utilitarian approach was followed to calculate the score; the set was given a one if the answer was both, half if either, and zero if none.\par 
Results in Table~\ref{tab:representativity-me} show a high correlation between the results obtained through the crowdsourcing method and the Persona with the human annotation method of $0.82$ Spearman Correlation, indicating the ability to substitute the model-based approach for the human-based approach in this context. However, as most of the participants were women, and jobs for the Middle Eastern race, except cashier, were non-inclusive of women, we can notice that the results of the crowdsourcing method are lower.

\begin{table}[]
 \centering
\caption{inclusion of representativity attributes scores of Middle Eastern generated images}
\label{tab:representativity-me}
\begin{tabular}{@{}lcc@{}}
\toprule
Occupation           & Crowdsourcing & Persona with Human annotation \\ \midrule
Baker                & 0.18             & 0.46           \\
Cashier              & 0.62              & 0.83           \\
Doctor               & 0.22             & 0.47          \\
Lawyer               & 0.12              & 0.46           \\
Secretary            & 0.27             & 0.62           \\
Software Developer   & 0.26                & 0.46           \\ \midrule
Pearson Correlation & \multicolumn{2}{c}{0.94}         \\ \midrule
Spearman Correlation & \multicolumn{2}{c}{0.82}         \\ \bottomrule
\end{tabular}
\end{table}



\begin{figure}
     \centering
          \begin{subfigure}[b]{0.45\textwidth}
         \centering
         \includegraphics[width=\textwidth]{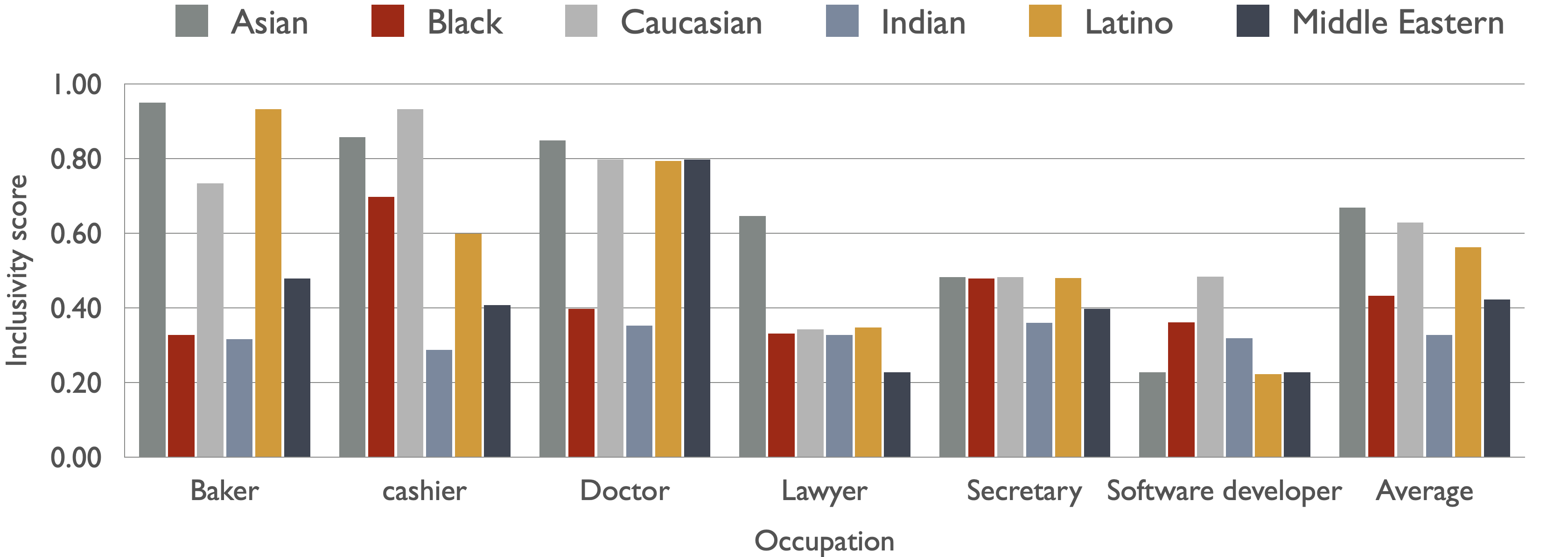}
         \caption{Model-based approach for calculating inclusion scores.}
     \end{subfigure}
     \begin{subfigure}[b]{0.45\textwidth}
         \centering
         \includegraphics[width=\textwidth]{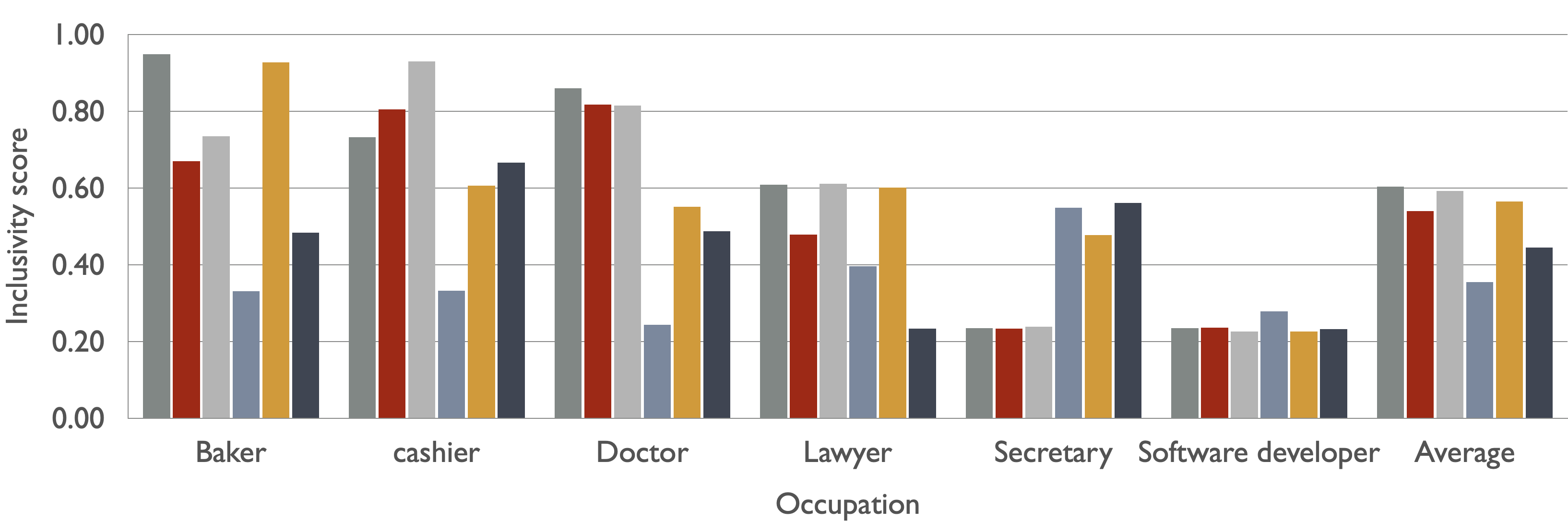}
         \caption{Human-based approach for calculating inclusion scores.}
     \end{subfigure}

     \begin{subfigure}[b]{0.45\textwidth}
         \centering
         \includegraphics[width=\textwidth]{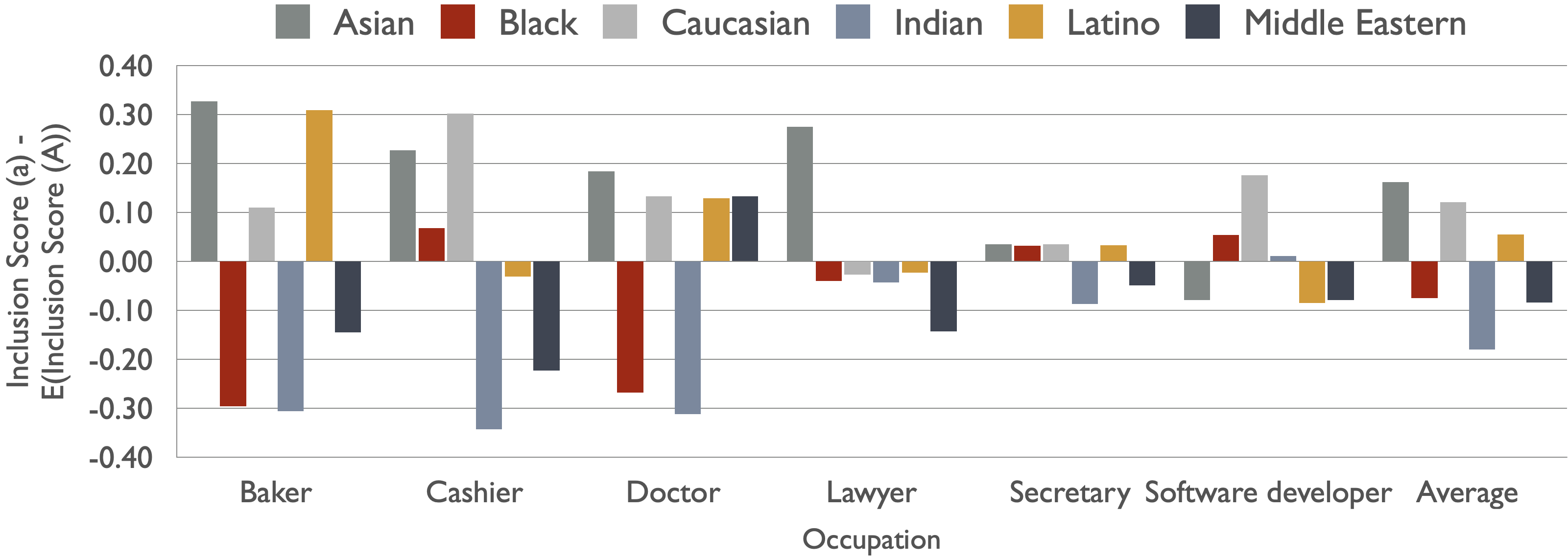}
         \caption{Score difference in the model-based approach.}
     \end{subfigure}
     \begin{subfigure}[b]{0.45\textwidth}
         \centering
         \includegraphics[width=\textwidth]{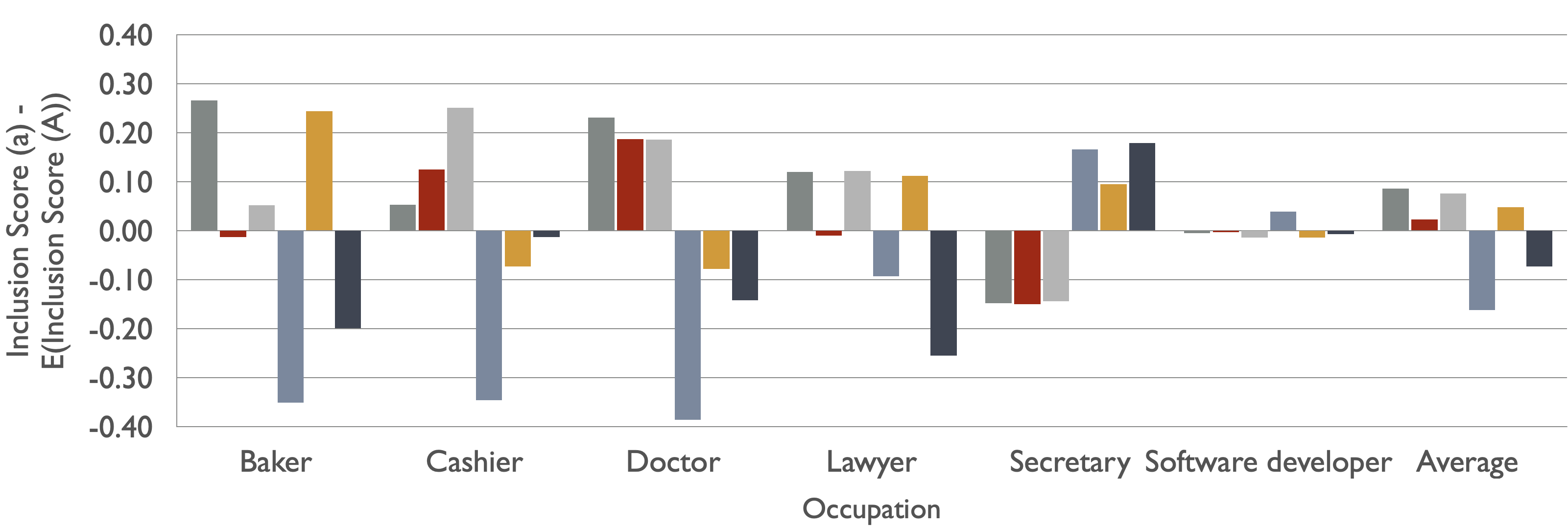}
         \caption{Score difference in the human-based approach.}
     \end{subfigure}

        \caption{Aggregated inclusion scores.}
        \label{fig:inc-scores-diff}
\end{figure}

\textbf{Aggregating Inclusion scores.} To produce the inclusion score, relevance scores and inclusion of representativity attributes scores are averaged per race/occupation combination. In the human-based approach, we averaged the scores of the single human evaluator for relevance and Persona with human annotation for representativity attributes scores. We averaged the scores from the image relevance pipeline and Persona with model annotation in the model-based approach. A Pearson correlation of 0.80 and a Spearman correlation 0f 0.79 was achieved, indicating a high correlation and that the approaches can be used interchangeably. Both approaches, as in Figure~\ref{fig:inc-scores-diff}, indicate lower inclusion scores for Middle Eastern and Indians. 

\subsubsection{Quality} The image quality here concerns how photorealistic the generated image is. Three approaches were proposed: annotation by a single annotator, crowdsourcing, and a quality classification model. \par
\emph{Human-based approach} A single evaluator of middle eastern origin performed the annotation. The evaluation at the start was from 1-3, where one is extremely deformed, two is photo-realistic with some deformation, and three is a photo-realistic image.
\begin{figure}
 \centering
 \includegraphics[width=0.7\textwidth]{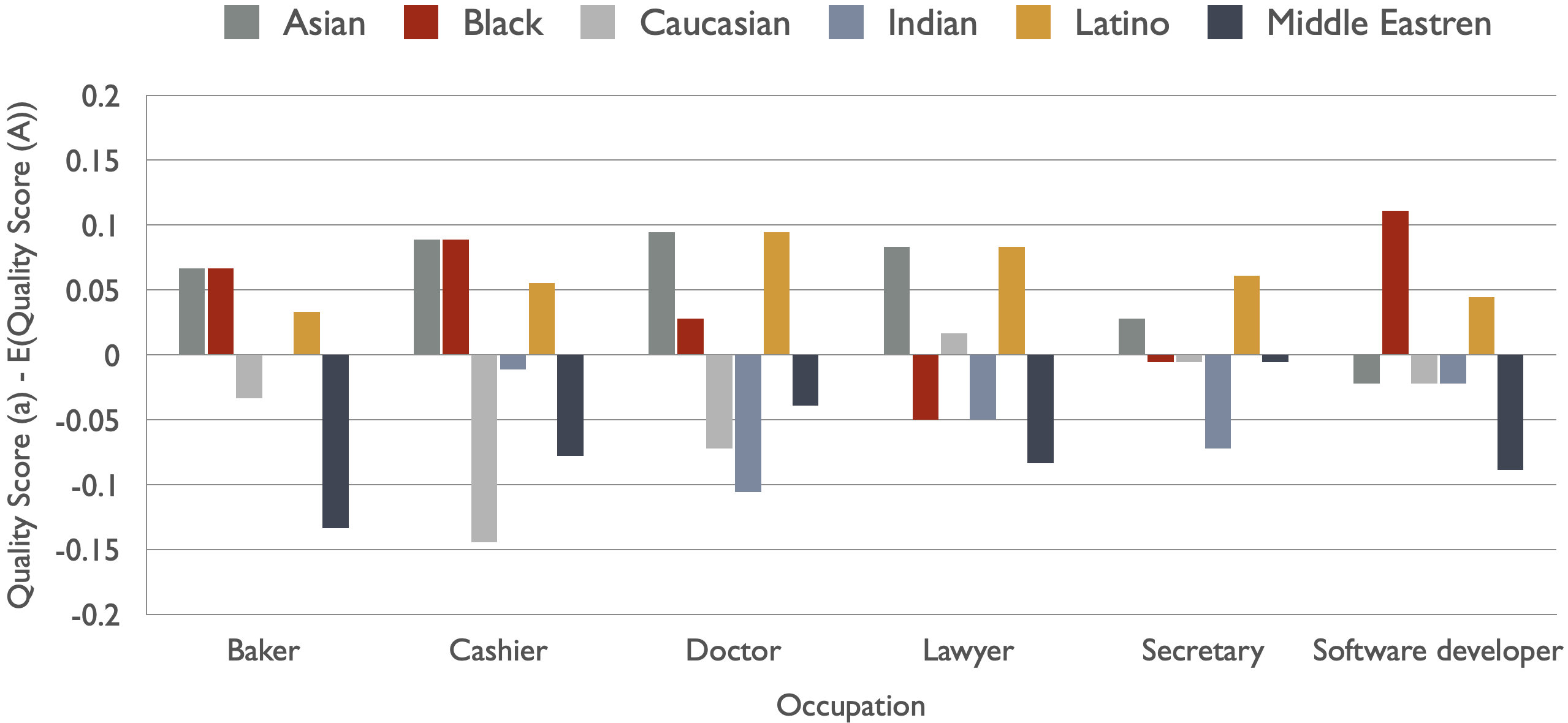}
\caption{Differences in Quality.}
\label{fig:quality}
\end{figure}

In another method for \emph{human-based approach}, via crowdsourcing, in the same questionnaire used to measure inclusion, the participants were asked to indicate the number of images they would use in a project per set. There was high conformance between the single human annotation and crowdsourcing when belonging to the same race, as shown in Figure~\ref{tab:quality-me}. However, for the different races, the pattern between the races as a whole is captured, but as an average across occupations. The average quality score is $2.48$ for the images generated for the South Asian/Indian race while crowdsourcing achieved $1.27$. On the other hand, images generated for the Black race achieved higher averages of $2.73$ and $2.60$ for the single human annotation and crowdsourcing approaches, respectively.

\begin{table}[]
\caption{Quality evaluation of Middle Eastern generated images}
\label{tab:quality-me}
\begin{tabular}{@{}lll@{}}
\toprule
Occupation           & Crowdsourcing & Single Annotator \\ \midrule
Baker                & 2.55             & 2.6           \\
Cashier              & 2                & 2.2           \\
Doctor               & 3.10             & 2.6           \\
Lawyer               & 2                & 2.2           \\
Secretary            & 2.76             & 2.4           \\
Software Developer                  & 2                & 2.4           \\ \midrule
Person Correlation & \multicolumn{2}{c}{0.80} \\
Spearman Correlation & \multicolumn{2}{c}{0.76}         \\ \bottomrule
\end{tabular}
\end{table}

\begin{figure}
 \centering
 \includegraphics[width=0.7\textwidth]{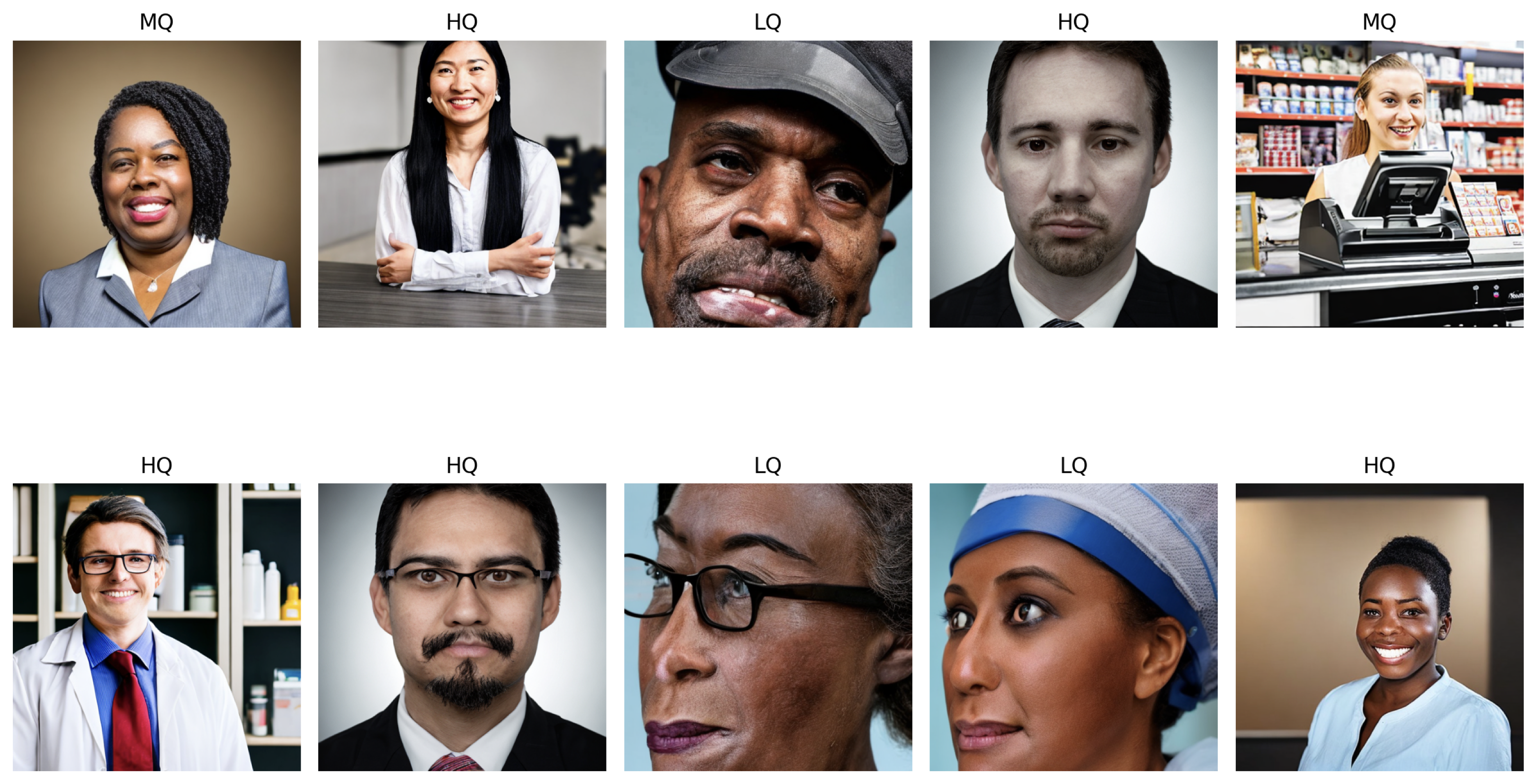}
\caption{Differences in Quality.}
\label{fig:gen_quality}
\end{figure}

\begin{figure}
     \centering
     \begin{subfigure}[b]{0.4\textwidth}
         \centering
         \includegraphics[width=\textwidth]{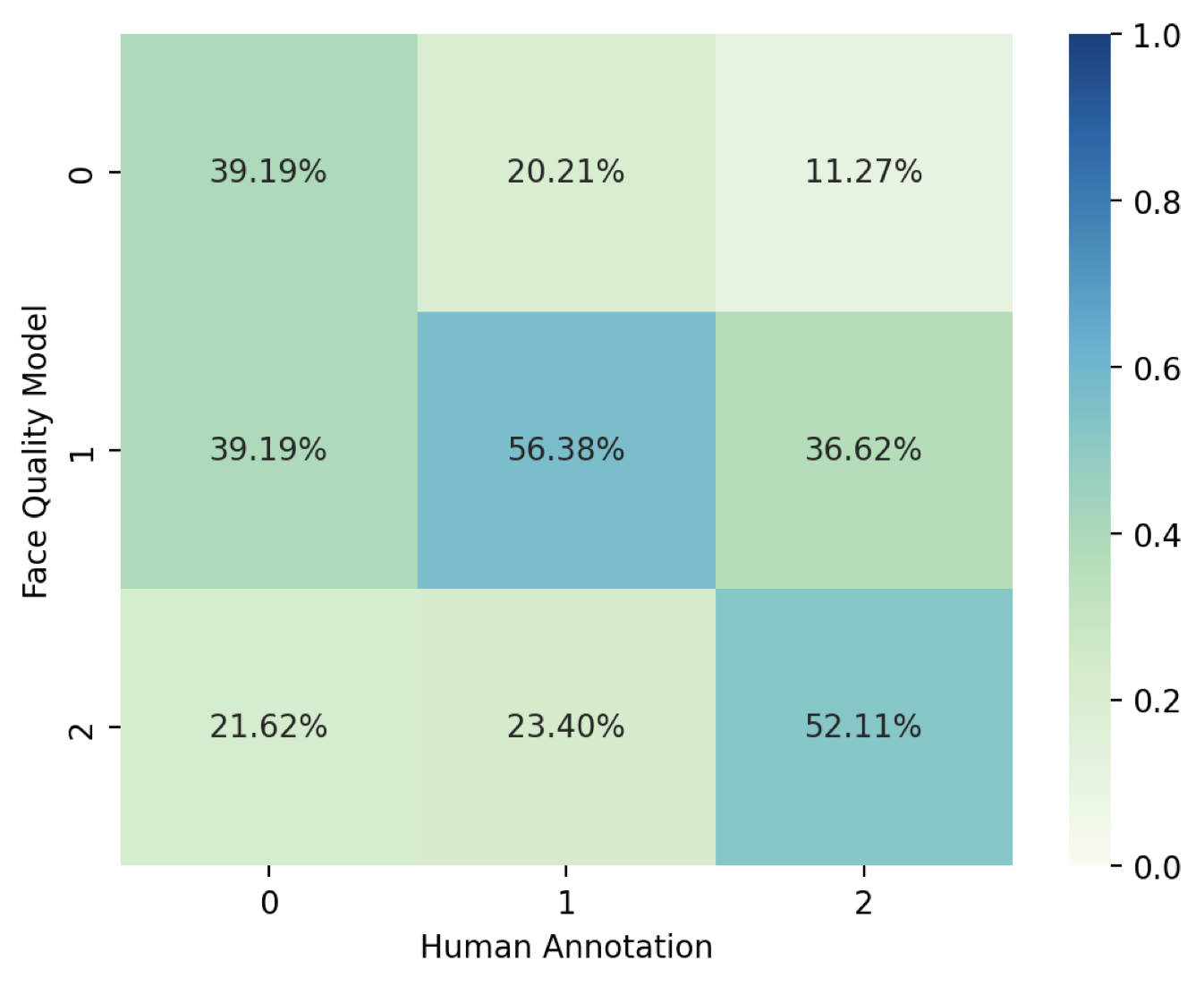}
         \caption{test-dev set}
     \end{subfigure}
     \begin{subfigure}[b]{0.4\textwidth}
         \centering
         \includegraphics[width=\textwidth]{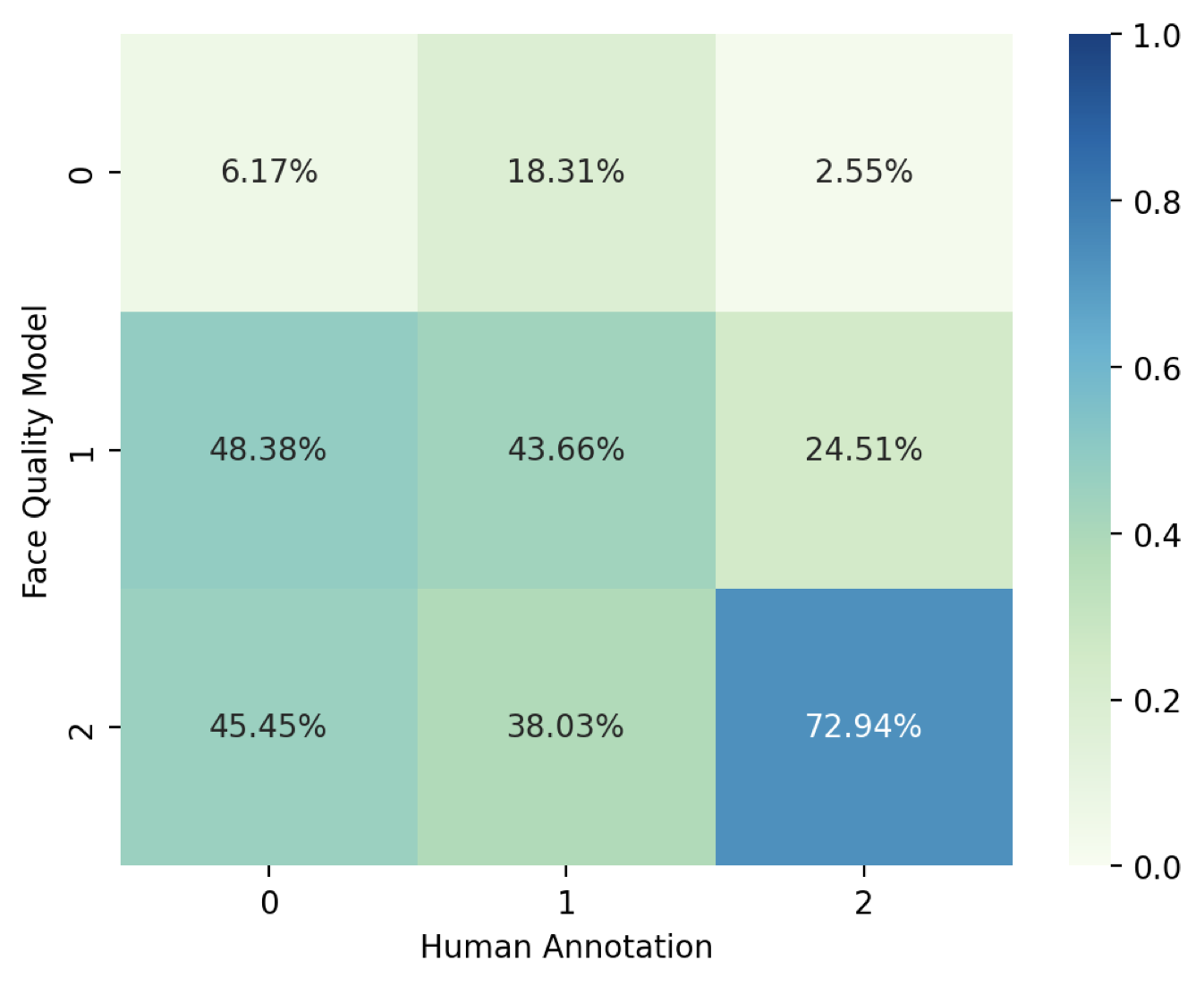}
         \caption{test set}
     \end{subfigure}
        \caption{Confusion matrix of the quality classification model.}
        \label{fig:conf_q}
\end{figure}

\emph{Model-based approach.} A ResNet50 \emph{Quality Classification Model} was finetuned on $1350$ generated images using different seeds and annotated for quality. A sample is in Figure~\ref{fig:gen_quality}. The model reached a $0.49$ macro average and weighted average f1-score on the test-dev subset. However, at testing the model on the $1110$ images under study, we can see the performance is reduced to $0.39$ macro average f1-score while maintaining a $0.50$ weighted average f1-score. The confusion matrices of the model's performance are shown in Figure~\ref{fig:conf_q}; zero maps to low quality, one to medium, and two to high quality.

\subsection{Decision} In the last step of the Framework, we answered the questions to decide on the presence of representativity bias in the studied model. Starting with diversity, the overall score is $0.65$ based on human evaluation and the KL-divergence metric. This is below the threshold of $0.70$. The Framework should stop here, indicating that the model carries representation bias. However, for illustration, we will proceed with the rest of the questions.
The next question answers if the images are inclusive; looking at the overall inclusion scores, we can see that the model based on both approaches for evaluation, with a low threshold of $0.55$,  is inclusive only for the ethnicities of Asians, Caucasians, and Latinos. We can conclude here that the model carries representational bias. However, proceeding to the last question, if the model holds multi-class statistical parity, in terms of quality, yes, it is fair. Although overall Middle Eastern-generated images have poorer image quality, it is within the threshold. As for inclusion multi-class statistical parity, No, it is not fair. The model is clearly biased against the Indian race. \par
\emph{With the dissatisfaction of the three criteria, the model exibit representativity bias in the context of occupation with respect to the sensitive attribute of race.
}
\section{Discussion} \label{sec:des}
Our proposed framework uses model-based and human-based approaches to evaluate the representativity fairness of TTI systems. It is able to capture the bias in the aspects of diversity, inclusion, and quality. The approaches show a high correlation in three out of four modules, validating the ability to substitute each other based on context and resources. In this section, we will discuss the proposed framework and its results by providing some recommendations, limitations, and threats to validity.  

\subsection{Recommendations}
\emph{Selecting thresholds and distribution:} Setting thresholds is highly dependent on the domain requirements and the laws of the regulators, if any. It also depends on the granularity level of $A$ and $Inc_{F}$. In our example, we could have reached the threshold for diversity if we only considered the binary values of white and non-white. Similarly, using the first and third prompts that included the phrase "A photo face of," we would have yielded higher relevance scores. Therefore, setting those thresholds is a matter of design and trial in the absence of regulations. In any case, we would not recommend setting thresholds below $0.65$ for diversity, $0.55$ for inclusion, and $0.20$ for the $\epsilon$ of multi-class statistical parity to avoid defying the purpose of the evaluation. Similarly, when selecting the distribution to measure the diversity against or generate personas, it should not be more biased than the reality. For example, changing the distribution from uniform to normal when selecting the persona's age yields a higher inclusion evaluation (around $0.02$ enhancement) as generated images are mostly in the age of $25-40$.  \par

\emph{Substituting human-based with model-based evaluation:} In three of the four modules designed for the evaluation, using a model-based evaluation would suffice with Spearman's correlations of $0.94$, $0.70$, and $0.82$ for diversity, relevance, and inclusion of representativity attributes scores, respectively. Of course, this is subject to the attributes chosen for diversity and inclusion of representativity attributes, as our approach covers race, age, and gender only in the context of occupation. We highly recommend validating the model-based approach for the attributes and context under study using a small subset of the generated images to issue any correction needed. As for quality, a single human annotator that belongs to each of the $A$ values will be needed at a minimum to minimize the annotation cost while maintaining a fair evaluation. It is worth noting that quality issues, if present for the context or sensitive attribute, do appear repetitively, so a subset of the generated images would be sufficient. 

\begin{figure}
 \centering
 \includegraphics[width=0.7\textwidth]{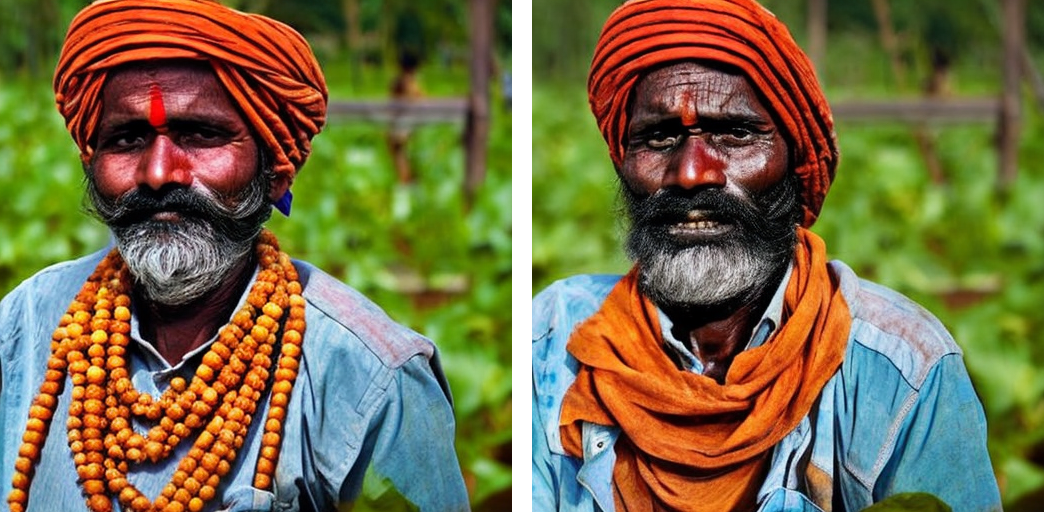}
\caption{Removing a bindi from an Indian farmer using semantic guidance.}
\label{fig:bindi}
\end{figure}

\emph{Guidance or Further Training:} Although using semantic guidance work in many cases related to gender~\cite{friedrich2023fair}, the model appears to lack learning an inclusive representation when it comes to stereotypes in race. Attempting to remove a 'bindi' from an Indian or a 'hijab' from a Middle Eastern  distorts the image. It invokes other stereotypes, such as more wrinkles and grey hair, as in Figure~\ref{fig:bindi}, or changes the gender or race, although both are explicit in the prompt. Another observation is that most Middle Eastern images face features look-alike, which makes you wonder; is guidance sufficient to steer the direction of the desired embedding in the latent space, or has the model not learned an inclusive embedding yet? Therefore, to avoid propagating bias in your application, we recommend experimenting with semantic guidance approaches to determine if they are sufficient for the members of the sensitive attribute using your tool.

\subsection{Limitations}
\begin{figure}
 \centering
 \includegraphics[width=0.7\textwidth]{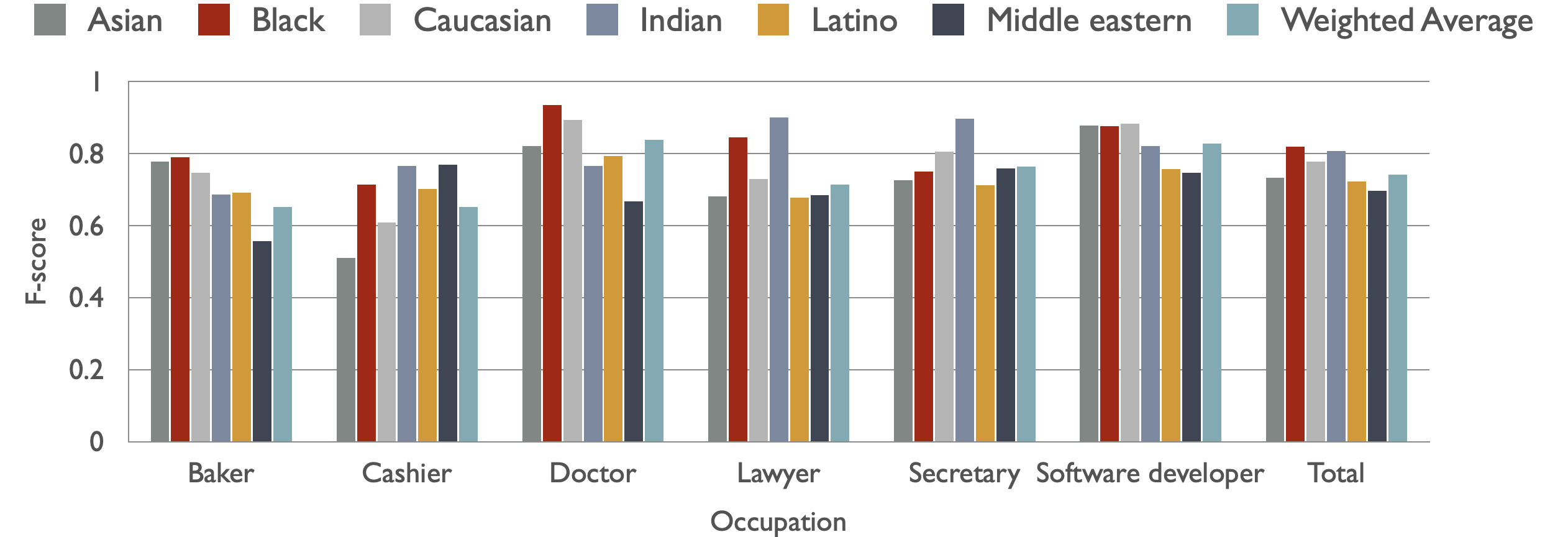}
\caption{f1-score per race and occupation.}
\label{fig:conf_diveresity}
\end{figure}

\emph{Limitations of the face analysis models with generative images:} The main cause of the minor discrepancies between the model and human annotation was the face analysis model deepFace. Starting with race, we investigate the f-score of the race classification across the different jobs based on all $1110$ images. Figure~\ref{fig:conf_diveresity} shows that deepFace has a lower f-score for the Middle Eastern, Latino, and Asian races. As for recall and precision, in $4$ out of $6$ occupation, the White/Caucasian race has a higher recall than precision, indicating the bias towards classifying a White/Caucasian race. As for the Middle Eastern, Latino, and Asian races, they are frequently confused with each other or misclassified as White. The deepFace face analysis model also has issues classifying the race grayscale-generated images. 

\begin{figure}
     \centering
     \begin{subfigure}[b]{0.4\textwidth}
         \centering
         \includegraphics[width=\textwidth]{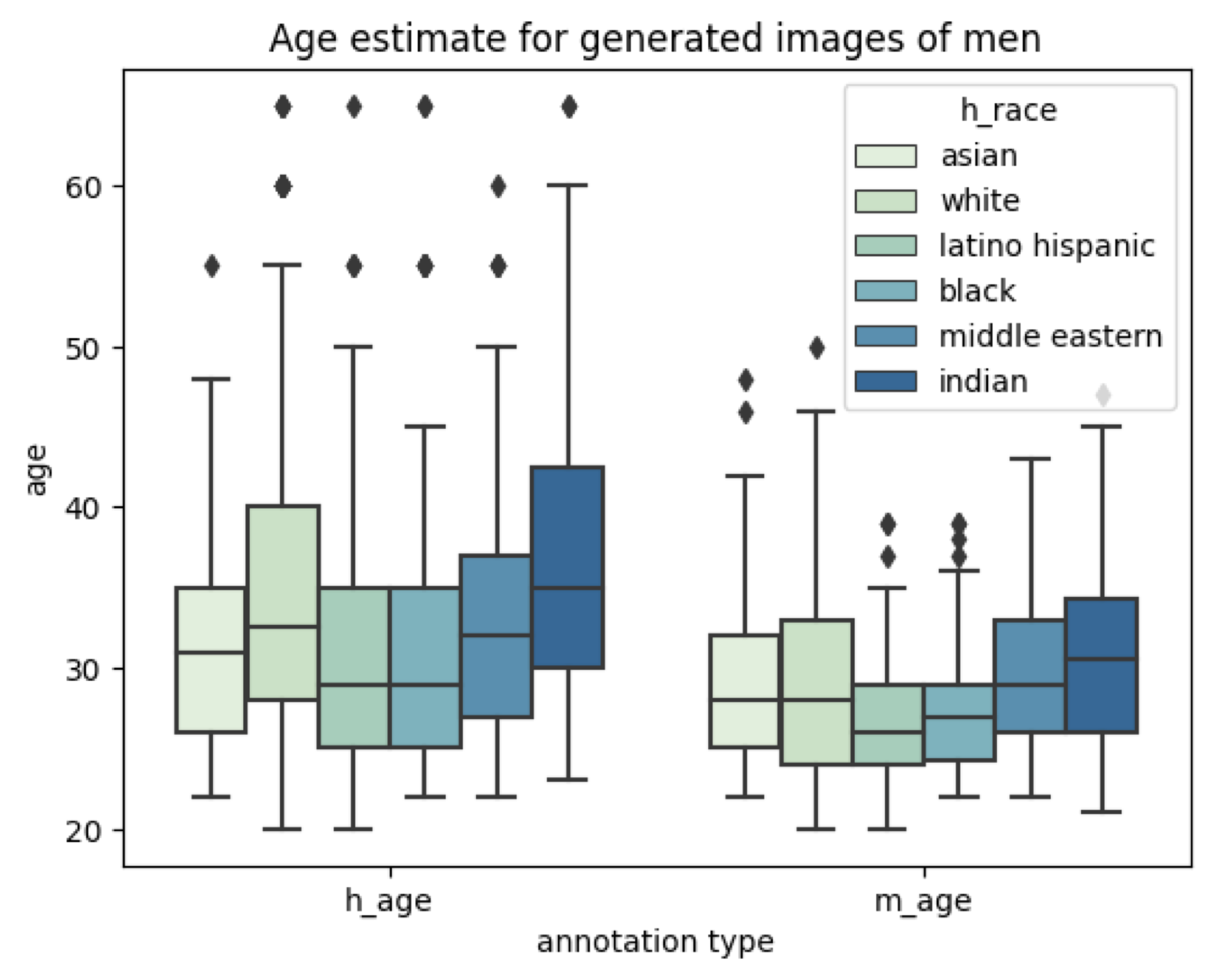}
         \subcaption{Men}
     \end{subfigure}
     \begin{subfigure}[b]{0.4\textwidth}
         \centering
         \includegraphics[width=\textwidth]{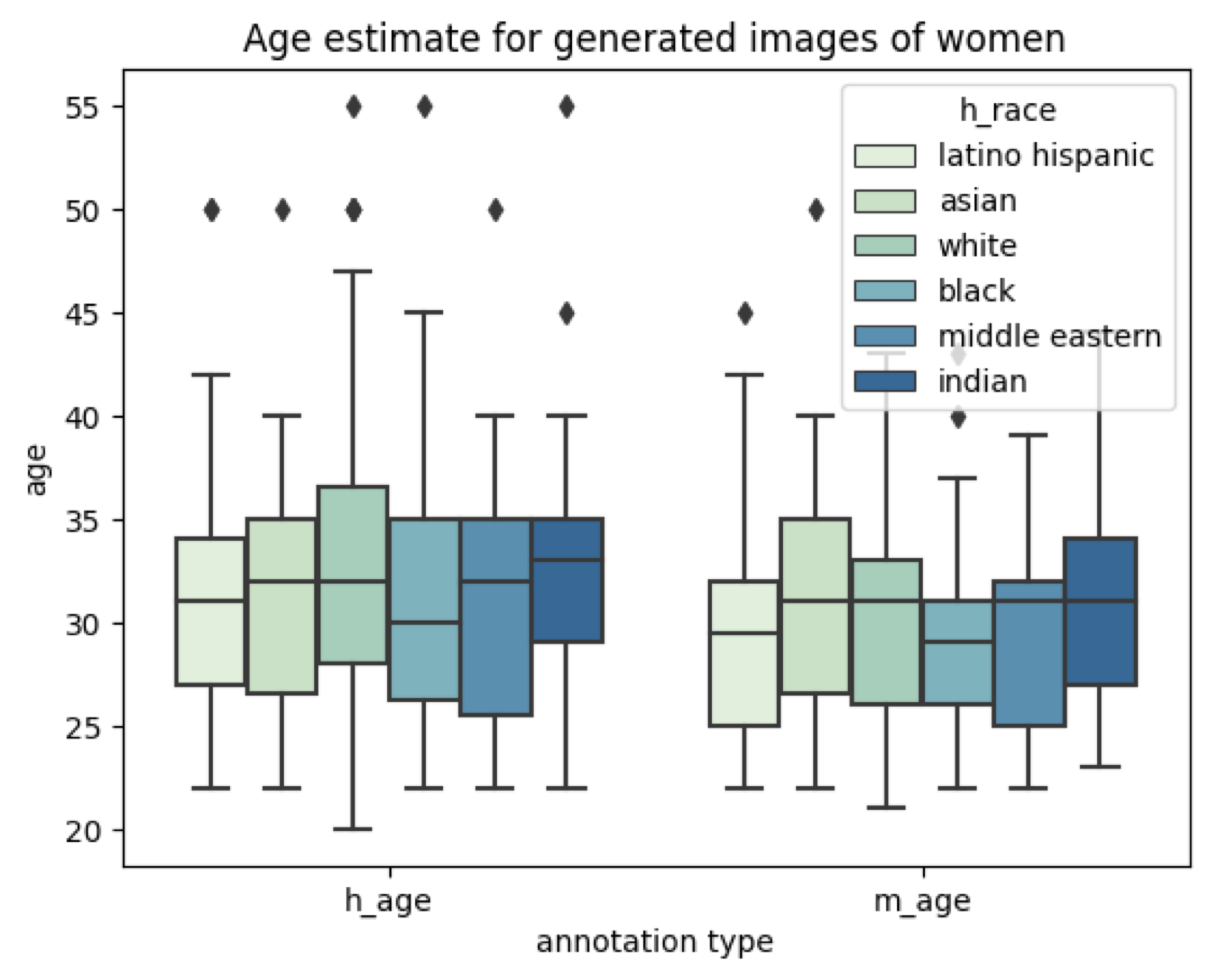}
                  \subcaption{Women}
     \end{subfigure}
        \caption{Age plot boxes: Human annotation (h\_age) has higher means and larger variance than deepFace annotation (m\_age). This observation is emphasized in generated faces of men. }
        \label{fig:conf-age}
\end{figure}

As for gender, when considering deepFace only, gender classification would drastically limit the Pearson correlation of inclusion of representativity attributes to $0.41$ as it has a $0.81$ average f1-score. However, when considering GIT first,  our gender classification approach has a $0.94$ averaged f1-score and a $0.80$ Pearson correlation. As for age bias, deepFace underestimates ages, especially for men, as illustrated in Figure~\ref{fig:conf-age}. 

\emph{Limitations of image relevance pipeline:} The pipeline's performance reduction mainly comes from the captioning module. It exhibits a lack of sensitivity to small backgrounds and blurred objects, producing a caption that does not capture the full context. For example, for baker-related images, "a woman in a chef's uniform stands behind a row of donuts." capture the background image, while "a man wearing a chef's apron stands in a kitchen" doesn't, although there were blurred bread loaves. The second issue is with the zero-shot classification model with the bias within the classes assigned. Although we did replace the man/woman with a person, the image-to-text model produced other gender markers such as beard and red lipstick, which does propagate to the classification model. For example, a man with a beard is less likely to be a secretary, while a woman with a beard is more likely to be a software developer. Therefore, it is essential to neutralize the output of the image-to-text model before feeding it to the classifier.\par

\emph{Limitations of model-based quality evaluation:}
Quality classification is the most challenging to substitute model evaluation for human evaluation. Several alternatives were considered for model-based evaluation, including the confidence score of detecting the ''person'' object of a YoloV8 model, which always detected the generated face as a  ''person'' with high confidence even when the face is deformed. AI image detectors \footnote{https://huggingface.co/umm-maybe/AI-image-detector} are almost always detecting the generated face as AI with a confidence score that does not correlate with the quality of the generated face. 
As for the classification model, although trained on around $3000$ labeled images, it still generalized poorly in Figure~\ref{fig:conf_q}. In addition, different versions of the model would require further finetuning of the classification model with many labeled images of the same version, which is unrealistic. The labels also change over time with respect to the big leaps in quality improvement across versions. \par

\emph{Limitations of human evaluation:} The diversity and representativity of the participants is important for the success of the crowdsourcing campaign. Otherwise, the evaluation could be far from the reality. It is also time-consuming to attempt annotating a large set of generated images. Therefore, reviewing the model-based annotation, although the model's output could influence it, would be the most efficient way to annotate a large set of images. In case of a sensitive context or a high budged, human review of the model's annotation is always recommended.

\subsection{Threats to  validity}
\emph{Threats to construct validity} concerns the degree to which the measures used represent what the researchers intended to look for. The most important threat is the aspects necessary to capture representativity fairness or bias. To mitigate this threat, we use both diversity and inclusivity, which are used repetitively to capture representativity concepts in both technical and social contexts. We also add quality multi-class statistical parity fairness to ensure demographical parity of the generated images. Another threat is the validity of metrics used in measuring those aspects. To mitigate this risk, we adopt metrics previously established in the literature in similar or analogous contexts.\par \emph{Threats to internal validity} concerns the annotation bias of the different attributes of the generated images and the selection bias of the models and thresholds. The generated images are of fictional characters; they do not hold the age, gender, and race attributes which are biological and social traits. Therefore, the annotations cannot be linked to ground truth and are based on the annotator's assumptions of the attributes. As the human annotations are also based on reviewing the model's annotation by the first author, they will tend to conform more to the model's annotation. To mitigate this risk, when a generated image does not show clear markers of an attribute, it was annotated with "-" and removed from the calculations. Also, the attributes were randomly reviewed by the second co-author. In addition, clear guidelines were set to annotate for quality and relevance for consistency. To mitigate the models' selection bias, we tried various models for each module in preliminary experiments to select the best fit. In addition, except for the quality classification model, all models considered are open-source and have wide adoption from ML practitioners. As thresholds do not have regulatory standards in the context under study, they were set by the first author and reviewed by the second author.\par \emph{Threats to external validity} concern the generalization of the study results. The study results are specific to Text-to-image generative systems and the occupation context, and further modifications and experiments must be conducted to be applied in other generative systems and contexts. Another threat is that most questionnaire respondents are from the middle east, and less than $5$ respondents for each race. To mitigate this threat, we only make conclusions on the correlation between the quality annotation and inclusion annotation when having a single annotator from the same race. We do not generalize the results to when having annotators from different races.

\section{Conclusion and Future work}
\label{sec:conc}
In this paper, we propose the Text-to-Image Representativity Fairness Evaluation Framework and demonstrate how it can evaluate TTI systems' diversity, inclusivity, and quality. Using human-based approaches and model-based approaches to capture such aspects, the framework provides alternatives that can substitute each other with high correlation based on the resources and context. The proposed framework  evaluated Stable Diffusion v2.1 in the context of occupation with respect to the sensitive attribute race and representitivity attributes gender and age. In the evaluation, models such as deepFace for face analysis and GIT for image captioning were used to capture diversity and inclusion of representativity attributes with $0.94$ and $0.82$ Spearman correlation, respectively. Using the proposed image relevance pipeline consisting of a GIT model and a zero-shot classification model holds a $0.70$ Spearman correlation when measuring relevance as a part of measuring inclusion. As for quality, the results show that a single human annotator from each member of the sensitive attribute is needed to evaluate the generated images' quality. The evaluation of Stable Diffusion v2.1 through the proposed framework displays enormous representativity issues in the context of jobs, especially for the races of Indians and Middle Easterners. Through the evaluation of model-based approaches, we also investigate deepFace limitations for labeling generated images for race, age, and gender. \par
Future work could focus on automating quality measurement with a semi-supervised or unsupervised approach that would adapt over time. Moreover, expanding the study to cover other contexts or include attributes such as skin tones, clothing objects, and emotions. Also, continual learning for these TTI systems enhances the learned representation of the disadvantaged demographics.

\printcredits

\bibliographystyle{model1-num-names}

\bibliography{main}


12

\end{document}